\documentclass[fleqn,usenatbib]{mnras}

\usepackage{newtxtext,newtxmath}
\usepackage[T1]{fontenc}

\DeclareRobustCommand{\VAN}[3]{#2}
\let\VANthebibliography\thebibliography
\def\thebibliography{\DeclareRobustCommand{\VAN}[3]{##3}\VANthebibliography}

\usepackage{graphicx}	% Including figure files
\usepackage{amsmath}	% Advanced maths commands
\usepackage{amssymb}	% Extra maths symbols
\usepackage{ulem}
\usepackage{subfigure}

\DeclareMathOperator\sech{sech}

\newcommand{\mrm}[1]{\mathrm{#1}}
\newcommand{\nuc}[2]{$\mrm{^{#2}#1}$}

\usepackage{xcolor}
\definecolor{purple2}{rgb}{0.5, 0.0, 0.5}

\title[Galactic 511\,keV Morphology]{Measuring the smearing of the Galactic 511\,keV signal: positron propagation or supernova kicks?}

\author[T. Siegert et al.]{
	Thomas Siegert,$^{1}$\thanks{E-mail: thomas.siegert@uni-wuerzburg.de}
	Roland M. Crocker,$^{2}$
	Oscar Macias,$^{3,4}$
	Fiona H. Panther,$^{5,6}$
	Francesca Calore,$^{7}$
	\newauthor
	~Deheng Song,$^{8}$
	and Shunsaku Horiuchi$^{8,3}$
	\\
	$^{1}$ 	Institut f\"ur Theoretische Physik und Astrophysik, Universit\"at W\"urzburg, Campus Hubland Nord, Emil-Fischer-Str. 31, 97074 W\"urzburg, Germany \\
	$^{2}$ Research School of Astronomy and Astrophysics, Australian National University, Canberra 2611, A.C.T., Australia \\
	$^{3}$ Kavli IPMU (WPI), UTIAS, The University of Tokyo, Kashiwa, Chiba 277-8583, Japann \\
	$^{4}$ GRAPPA Institute, University of Amsterdam, 1098 XH Amsterdam, The Netherlands \\
	$^{5}$ Department of Physics, University of Western Australia, Crawley WA 6009, Australia\\
	$^{6}$ Australian Research Council Centre of Excellence for Gravitational Wave Discovery (OzGrav) \\
	$^{7}$ University of Grenoble Alpes, Université Savoie Mont Blanc, Centre national de la recherche scientifique, \\ ~~~Laboratoire d'Annecy-le-Vieux de Physique Théorique, Annecy, France \\
	$^{8}$ Center for Neutrino Physics, Department of Physics, Virginia Tech, Blacksburg, VA, USA 
}

\date{Accepted XXX. Received YYY; in original form ZZZ}

\pubyear{2021}

\begin{document}
	\label{firstpage}
	\pagerange{\pageref{firstpage}--\pageref{lastpage}}
	\maketitle
	
	\begin{abstract}
		\noindent %The 511\,keV signal from the annihilation of positrons with electrons in the Milky Way has presented a puzzle for half a century.
		We use 15 years of $\gamma$-ray data from INTEGRAL/SPI in a refined investigation of the morphology of the Galactic bulge positron annihilation signal.
		Our spatial analysis confirms that the signal traces the old stellar population in the bulge and reveals for the first time that it traces the 
		boxy bulge and nuclear stellar bulge.
		Using a 3D smoothing kernel, we find that the signal is smeared out over a characteristic length scale of $150 \pm 50$\,pc, suggesting either annihilation \textit{in situ} at astrophysical sources kicked at formation or positron propagation away from sources.
		The former is disfavoured by its requiring kick velocities different between the Galactic nucleus ($\gtrsim 50\,\mrm{km\,s^{-1}}$) and  wider bulge ($\lesssim 15\,\mrm{km\,s^{-1}}$) source.
		Positron propagation prior to annihilation can explain the overall phenomenology of the 511\,keV signal for positrons injection energies $\lesssim 1.4$\,MeV, suggesting a nucleosynthesis origin.
	\end{abstract}

	
	%%%%%%%%%%%%%%%%%%%%%%%%%%%%%%%%%%%%%%%%%%%%%%%%%%
	
	%%%%%%%%%%%%%%%%% BODY OF PAPER %%%%%%%%%%%%%%%%%%
	
	\section{Introduction}\label{sec:intro}
	The strongest persistent, diffuse $\gamma$-ray line signal is found at 511\,keV photon energies and originates from the annihilation of positrons with electrons.
	This signal was first detected by a balloon-borne instrument in 1969 \citep{Johnson1973_511}.
	The line's	flux, of the order of $10^{-3}\,\mrm{ph\,cm^{-2}\,s^{-1}}$, indicates an annihilation rate of $10^{50}\,\mrm{e^+\,yr^{-1}}$, with a high concentration in the Galactic bulge \citep{Purcell1997_511,Knoedlseder2005_511}, but the origins of the vast amount of antimatter implied are still unclear \citep{Prantzos2011_511}.
	Theoretical expectations associate astrophysical positron sources with regions hosting on-going star formation in the Galactic disc \citep[e.g.,][]{Alexis2014_511ISM}, leading to an expected bulge-to-disc luminosity ratio of $(B/D)_{511} \lesssim 0.5$.
	This expectation has been enduringly confounded by measurements, the most recent of which point to $(B/D)_{511} \sim 1$ \citep{Siegert2016_511}.
	%\footnote{Note that earlier measurements \citep{Knoedlseder2005_511,Weidenspointner2008_511,Bouchet2010_511} pointed to even higher ratios $\sim$\,3--9.}.
	%
	This anomaly has been interpreted in two ways: 
	Either positrons propagate away from their putative disc sources, becoming subsequently trapped in the Galactic bulge \citep[e.g.,][]{Prantzos2006_511,Higdon2009_511}, or a central source populates the bulge by positron transport, either diffusive \citep{Jean2009_511ISM,Martin2012_511,Alexis2014_511ISM} or advective \citep{Jean2009_511ISM,Crocker2011_FB,Panther2018_nuclear_outflow,Churazov2011_511}.
	These interpretations, however, fail to reproduce details in the morphology or violate kinematic constraints.
	Both scenarios require the transport of positrons over $\mathrm{kpc}$ distances. 
	The possibility that a large fraction of positrons might annihilate either \textit{in situ} at their sources has attracted little attention \citep[e.g.,][]{Martin2010_SN511}.
	One known but subdominant positron source is the radioactive isotope \nuc{Al}{26} generated in massive star nucleosynthesis.
	\nuc{Al}{26} nuclei create positrons at $\lesssim\mrm{MeV}$ energies, hundreds of parsecs away from the massive stars, implying that the 511\,keV emission will not exactly trace \nuc{Al}{26} sources.
	A generalisation of this scenario to other candidate sources is \textit{prima facie} plausible, but has not previously found supporting evidence.

	Given positrons might annihilate at any distance from their source(s), here we attempt to constrain their propagation length by a spatial analysis of the 511\,keV line and the ortho-Positronium (ortho-Ps) continuum in the Milky Way bulge from 15 years of INTEGRAL/SPI data.
	\citet{Weidenspointner2006_oPs} reconstructed an image in the energy band 410--500\,keV where ortho-Ps is dominant, and found an asymmetric bulge component.
	Here we perform for the first time a spectrally resolved study
	of ortho-Ps that extends to lower energies, 200--500\,keV.
	This leads to an improvement in statistical quality, and boosts the discriminant power of existing data.
	Including ortho-Ps allows us to compare various spatial tracers of positron annihilation in the Galaxy by the use of stellar templates, dark matter profiles, or discy gas maps.
	We determine that the emission is blurred when compared to the best-fitting tracer, and show that this is explained by positron propagation rather than supernova kicks.

	\section{Positron Sources in the Milky Way}\label{sec:sources}
	Among the most promising sources of positrons are $\beta^+$-unstable nucleosynthesis products from thermonuclear supernovae \citep[\nuc{Co}{56}, \nuc{Ti}{44}; e.g.,][]{Milne1997_511,Diehl2015_SN2014J_Co,Churazov2015_2014JCo,Crocker2017_511_91bg}, massive stars \citep[\nuc{Al}{26}; e.g.,][]{Mahoney1984_26Al,Oberlack1996_26Al,Diehl2006_26Al,Kretschmer2013_26Al,Pleintinger2019_26Al}, and core-collapse supernovae \citep[\nuc{Ti}{44}; e.g.,][]{Iyudin1997_CasA,Grebenev2012_SN1987A,Siegert2015_CasA,Boggs2015_SN1987A,Grefenstette2014_CasA,Grefenstette2017_CasA,Weinberger2020_CasA}.
	Additional, putative sources include compact objects \citep[e.g.,][]{Bouchet1991_mq,Sunyaev1992_xrb511,Weidenspointner2008_511,Siegert2016_V404,Bartels2018_binaries511}, such as accreting black holes and magnetised neutron stars, and also dark matter \citep[DM; e.g.,][]{Boehm2004_dm,Finkbeiner2007_dm511deex,Siegert2016_dsph}.
	Positrons are also known products in solar flares where they annihilate in the Sun's atmosphere \citep{Murphy2005_posiloss}, constituting a special case of \textit{in-situ} annihilation.
	Thus, the population of intermittently flaring stars could partly explain the measurements \citep{Bisnovatyi-Kogan2017_511}.
	The stellar population in the bulge and centre of the Galaxy is seen at infrared wavelengths, so that in-situ annihilation may coincide with models describing the nuclear stellar bulge \citep[NB;][]{Launhardt2002_NB} and larger boxy bulge \citep[BB;][]{Freudenreich1998_BoxyBulge_COBE}.

	The fractional contribution of each of these source populations has not yet been determined and the identity of a dominant positron source -- if such exists -- remains subject to debate.
	A measurement of the characteristic length scale over which positrons propagate through the interstellar medium (ISM) before annihilating may resolve this impasse.
	This follows from the fact that different source types are characterised by different injection energies, and this strongly determines the maximum distance that positrons can propagate.

	The transport of positrons in the ISM is dictated by the Galactic magnetic field which causes the positrons to scatter or stream along field lines.
	Their energy losses are determined by collisions with neutral (ionisation) and charged (Coulomb) particles \citep{Jean2009_511ISM,Martin2012_511,Alexis2014_511ISM,Panther2018_nuclear_outflow}. 
	Bremsstrahlung, Inverse Compton (IC), and synchrotron emission make subdominant contributions to positron cooling at MeV energies.
	Positrons in the ISM are expected to propagate until, upon encountering dense ($\gg 1\,\mrm{cm^{-3}}$) neutral material, they rapidly lose their remaining energy and annihilate.
	Heavier nuclei ($Z > 2$) may also be of importance when considering positrons that thermalise before annihilation \citep{Panther2018_alkali511}.
	When positrons manage to escape their source environments, the final annihilation regions are expected to be traced mainly by CO or HI emission.
	
	MeV $\gamma$-ray continuum observations \citep{Strong2005_gammaconti,Strong2010_CR_luminosity} constrain the injection energies of (annihilating) positrons to $\lesssim 3$--$7$\,MeV \citep{Sizun2006_511,Beacom2006_511}.
	The observed IC morphology provides constraints on where positrons end up thermalising and possibly annihilating `in flight' \citep[e.g.,][]{Aharonyan1981_511}.
	Instead of slowing down and annihilating, some positrons might also escape the Galaxy in a nuclear outflow associated with the Fermi Bubbles \citep[FB;][]{Su2010_fermibubbles,Crocker2011_FB}.

	\begin{table}
		\centering
		\caption{Summary of template maps used in this work. The `Built' column indicates observed or synthetic templates.}
		\begin{tabular}{lccc}
			\hline
			\hline
			Map & Process/Population                                  & Built               \\
			\hline
			\texttt{BB}  & stars ($1.25$--$4.9\,\mrm{\mu m})$     & obs. + syn.        \\
			\texttt{NB}  & stars  ($2.2$--$240\,\mrm{\mu m})$     & obs. + syn. \\
			\texttt{XB}  & stars ($3.4$--$4.6\,\mrm{\mu m})$     & obs. + syn. \\ 
			\texttt{IC}  & $\mrm{e^{\pm}+\gamma \rightarrow e^{\pm}+\gamma}$ & syn.        \\
			\texttt{CO}  & CO, $\mrm{J=1 \rightarrow 0}$ at 115\,GHz            & obs.        \\
			\texttt{HI}  & H, $\mrm{F=1 \rightarrow 0}$ at 21\,cm               & obs.        \\
			\texttt{DM0} & Dark matter, $\rho_{\rm NFW}^2(\gamma=1.0)$             & syn.        \\
			\texttt{DM2} & Dark matter, $\rho_{\rm NFW}^2(\gamma=1.2)$             & syn.        \\
			\texttt{FB}  & Fermi Bubbles (GeV)                                 & obs. + syn. \\
			\hline
			\hline
		\end{tabular}
		\label{tab:maps}
	\end{table}

	\section{Emission Models}\label{sec:emission_models}
	\subsection{Template Maps}\label{sec:3D_models}
	Tab.\,\ref{tab:maps} gives an overview of positron annihilation emission morphologies tested in this paper.
	The template maps may trace either populations associated with emission in the nominated wavelength bands or the emitting population itself.
	For example, a halo morphology can test for an old population of objects ($\gtrsim 1\,\mrm{Gyr}$), such as type Ia supernovae, or the stars of the halo themselves.

	If an analytic 3D-density distribution is available, we perform line-of-sight (los) integrations to determine the flux per steradian, $F(l,b)$, on a $(50^{\circ} \times 50^{\circ})$-sized grid of pixels with solid angles $(0.5^{\circ} \times 0.5^{\circ})$,
	\begin{equation}
		F(l,b) = \frac{1}{4\pi} \int_{\rm los} \rho(x(s),y(s),z(s)) ds\mrm{.}
		\label{eq:los_integral}
	\end{equation}
	\noindent Here, $x(s) = x_{\odot} - s \cos(l) \cos(b)$, $y(s) = y_{\odot} - \sin(l) \cos(b)$, and $z(s) = z_{\odot} - s \sin(b)$ is the los vector along Galactic coordinates $(l,b)$.
	Details on the emission templates are found in appendix\,\ref{sec:emission_templates}.

	We test for bulge populations using a BB template \citep{Freudenreich1998_BoxyBulge_COBE}, an X-shaped Bulge \citep[XB;][]{Ness2016_Xbulge_WISE}, and the NB \citep{Launhardt2002_NB} population as was used by earlier Fermi-LAT GeV analysis \cite[e.g.,][]{Macias2018_LATGeV,Bartels2018_GeVexcess_stars}.
	Disc contributions are tested via the Planck CO map \citep{Planck2016_foregrounds}, an los-integrated HI map from 21\,cm radio observations \citep{Dickey1990_HI}, and  energy-dependent IC scattering template maps derived using GALPROP  \citep{Strong2007_GALPROP}.
	The halo is modelled either as a Navarro-Frenk-White \citep{Navarro1997_NFW} DM halo profile with a slope $\gamma = 1.0$ (DM0) or $\gamma=1.2$ (DM2), or an isotropic emission model with the shape of the FBs \citep{Su2010_fermibubbles}.

	\subsection{3D Blurring of Density Profiles}\label{sec:blurring_maps}
	For assessing how similar the initial templates are to the actually measured positron annihilation signal, we perform an additional analysis on the best-fitting images (see Sec.\,\ref{sec:results}).
	Blurred density profiles $\tilde{\rho}$,
	\begin{equation}
		\tilde{\rho}(x,y,z;\xi) = \rho(x,y,z) \otimes G(x,y,z;\Sigma)\mrm{,} 
		\label{eq:3Dblurring}
	\end{equation}
	\noindent are calculated by convolution ($\otimes$) with a 3D-Gaussian  $G(x,y,z;\Sigma)$ with a diagonal covariance matrix $\Sigma = \mathbb{I}\xi^2$ so that each spatial dimension obtains the same blurring length scale $\xi$.
	The resulting 3D-arrays of $\tilde{\rho}$ are then integrated along each los by trilinear interpolation ($(x,y,z) \rightarrow (l,b,s)$).
	We choose a 3D-grid in $x$ and $y$ between $\pm 6$\,kpc, and $z$ between $\pm 4$\, kpc in 50\,pc steps, and define a set of $\xi$-values between 0 and 600\,pc in 25\,pc steps for fits to the data.

	\section{INTEGRAL/SPI data analysis}\label{sec:data_analysis}
	\subsection{Data Set}\label{sec:data_set}
	ESA's INTEGRAL satellite \citep{Winkler2003_INTEGRAL} with its coded-mask $\gamma$-ray spectrometer SPI \citep{Vedrenne2003_SPI} has surveyed the sky since 2002. 
	Here we make use of SPI data collected between March 2003 and September 2017, focusing on the Galactic bulge, with a homogeneous exposure of 24.3\,Ms in the central $15^{\circ} \times 15^{\circ}$.
	This results in exposures out to $\lesssim 25^{\circ}$ in $l$ and $b$, given SPI's large field of view of $16^{\circ} \times 16^{\circ}$. 
	After removing outliers in the data set, it comprises 12587 pointings (targeted observations), with an average observation time of $2313$\,s, for a total on-time of $29.1$\,Ms.
	Up to 2021, four of 19 SPI detectors failed, reducing the sensitivity by $\approx 20\,\%$.

	We choose six logarithmic energy bands between 189 and 1805\,keV, plus a 6\,keV broad bin ($508$--$514$\,keV) for the 511\,keV line, to perform a bin-by-bin analysis.
	In this way we are able to capture dominant emission features belonging to positron annihilation, diffuse $\gamma$-ray continuum, and point sources (see appendix\,\ref{sec:data_set_details}).
	
	\begin{figure}
		\centering
		\includegraphics[width=0.8\columnwidth,trim=0.10in 0.7in 0.94in 0.83in,clip=true]{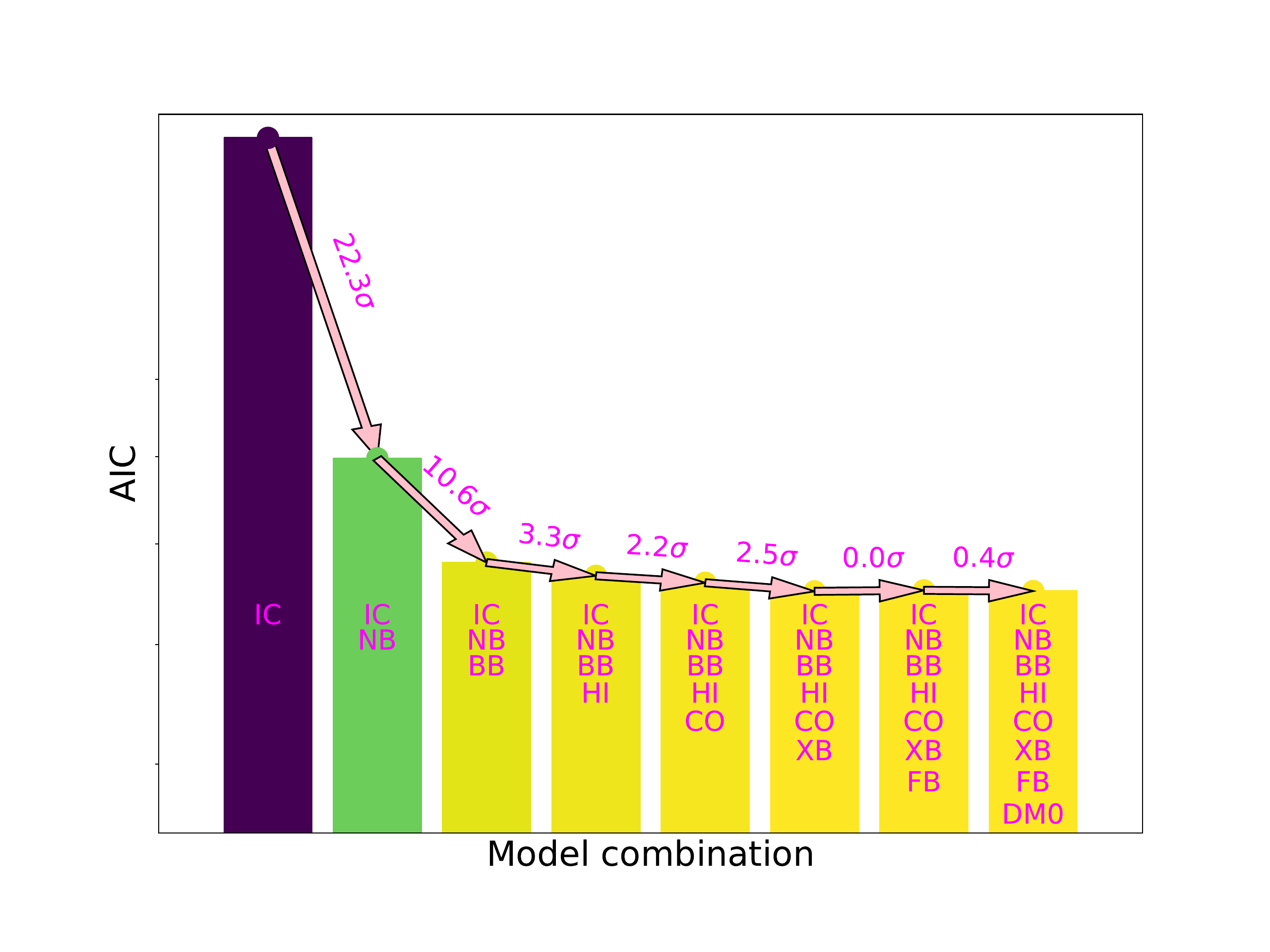}%
		\caption{Fit improvements when adding more templates (from left to right) for the energy band 189--1805\,keV. Relative AIC values convert to $\sigma$-values.}
		\label{fig:model_improvement_all_models}%
	\end{figure}

	\subsection{General Method}\label{sec:likelihood_analysis}
	SPI data analysis relies on the comparison of models, e.g. images or point sources, and a description of the instrumental background, to the raw photon count data.
	The chosen images are convolved through SPI's coded-mask response, which depends on the sources' aspect angles and the photon energy.
	The background model utilises knowledge about the instrument, its detectors and shields, and the long-term behaviour in space \citep{Diehl2018_BGRDB,Siegert2019_SPIBG}.
	
	The models are realised as time-series of expected detector illuminations (patterns) which are fitted to the measured time-series via a maximum likelihood fit using the Poisson likelihood
	\begin{equation}
		\mathcal{L}(d_p|\theta_i) = \prod_p^{N_{\rm obs}} \frac{m_p^{d_p} \exp(-m_p)}{d_p!}\mrm{,}
		\label{eq:cstat}
	\end{equation}
	\noindent where $d_p$ are the measured data in pointing $p$ out of $N_{\rm obs}$ observations, $m_p$ is the model prediction, and $\theta_i$ are free parameters, determining the fluxes of model components $i$.
	This is done for each energy bin that we define in our analysis.
	Our model is separated into sky and instrumental background contributions,
	\begin{equation}
		m_p = \sum_{t} \sum_{j} R_{jp} \sum_{k=1}^{N_S} \theta_{k,t} M_{kj} + \sum_{t'} \sum_{k=N_S+1}^{N_S+N_B} \theta_{k,t'} B_{kp}\mrm{.}
		\label{eq:spimodfit_model}
	\end{equation}
	\noindent In Eq.\,(\ref{eq:spimodfit_model}), $M_{kj}$ is the k-th of $N_S$ sky models to which the response $R_{jp}$ is applied for each pointing $p$ and pixel $j$.
	The $N_B = 2$ background models $B_{kp}$ are independent of the response.
	While diffuse Galactic and positron-annihilation emission are constant over long periods of time \citep{Purcell1997_511}, individual sources can change their spectral behaviour on shorter timescales which we account for by introducing a point source variability time scale $t$.
	Likewise, the background amplitude can change on a different time scale, $t'$, which is determined individually for each energy bin \citep{Siegert2019_SPIBG}.

	\subsection{Model comparisons}\label{sec:model_comparisons}\label{sec:model_combinations}
	For model comparisons, we use the Akaike Information Criterion \citep[AIC;][]{Akaike1974_AIC,Burnham2004_AICBIC},
	\begin{equation}
		\mrm{AIC} = 2 N_{\rm par} - 2 \ln\mathcal{L}(d_p|\theta_i)\mrm{,}
		\label{eq:AICc}
	\end{equation}
	\noindent with $N_{\rm par}$ being the number of fitted parameters.
	We compute AIC differences $\Delta\mrm{AIC}$ to judge the relative importance of model templates compared to a chosen baseline model (Sec.\,\ref{sec:baseline_model}).
	Relative AIC values take into account the differences in the likelihood and the degrees of freedom ($\nu$) so that a model with more parameters can result in a `worse' fit.
	Improvements in terms of $\sigma$-values are estimated by equating the survival probability of a $\chi^2_{\nu}$-distribution to the relative likelihood $\exp\left(-\Delta\mrm{AIC}/2\right)$ \citep{Burnham2004_AICBIC}.
	Given our nine template emission models, this amounts to 512 different combinations for each of the seven energy bins.
	Because combinations including DM0+DM2 cannot be realised in nature, we exclude them in the following.
	In the first bin, 189--245\,keV, several point sources contribute to the total flux, but become negligible above 500\,keV.	
	As an example, in Fig.\,\ref{fig:model_improvement_all_models} we show the 
	gradual improvement in the fit one achieves for a particular concatenation of templates.
	We choose the optimum combination for positron annihilation, IC+NB+BB (see Sec.\,\ref{sec:maps_as_they_are}), as a starting template.
	The next component is chosen on the basis of which of the remaining provides the largest $\Delta\mrm{AIC}$ over the total energy band analysed ($189$--$1805$\,keV).
	We proceed in this fashion until the remaining templates are linked into the chain.

	\begin{table}
		\centering
		\caption{Summary of model comparisons. We show the AIC of additional source templates (second column) with respect to the baseline model for the 511\,keV line, the ortho-Ps bins (oPs; $245$--$508$\,keV) as well as positron annihilation as a whole. The second and third sections summarise the search for additional components above the best-fitting BB+NB stellar template.}
		\begin{tabular}{lcrrr}
			\hline
			\hline
			Baseline model & Add. source & $\Delta\mrm{AIC}_{511}$ & $\Delta\mrm{AIC}_{\rm oPs}$ & $\Delta\mrm{AIC}_{\pm}$ \\
			\hline
			\texttt{IC} & \texttt{HI} & 10.9 & 4.7 & 15.6  \\
			\texttt{IC} & \texttt{FB} & 25.2 & 9.9 & 35.1  \\
			\texttt{IC} & \texttt{BB} & 89.1 & 192.4 & 281.5  \\
			\texttt{IC} & \texttt{CO} & 64.6 & 239.0 & 303.6  \\
			\texttt{IC} & \texttt{HI+CO} & 104.5 & 278.1 & 382.6  \\
			\texttt{IC} & \texttt{XB} & 105.7 & 289.5 &  395.2  \\
			\texttt{IC} & \texttt{NB} & 123.8 & 383.8 & 507.6  \\
			\texttt{IC} & \texttt{DM2} & 134.8 & 375.8 & 510.6  \\
			\texttt{IC} & \texttt{DM0} & 164.3 & 433.3 & 597.6  \\
			\textbf{\texttt{IC}} & \textbf{\texttt{BB+NB}} & \textbf{162.0} & \textbf{456.2} & \textbf{618.2} \\
			\hline
			\texttt{IC+BB+NB} & \texttt{CO} & -2.0 & -1.7 & -3.7  \\
			%\texttt{IC+BB+NB} & \texttt{DM2} & -0.5 & -0.8 & -1.3  \\
			\texttt{IC+BB+NB} & \texttt{DM0} & 3.6 & -1.1 & 2.5  \\
			\texttt{IC+BB+NB} & \texttt{CO+HI} & -1.4 & 16.8 & 15.4  \\
			\textbf{\texttt{IC+BB+NB}} & \textbf{\texttt{HI}} & \textbf{-0.3} & \textbf{16.3} & \textbf{16.0}  \\
			\hline
			\texttt{IC+BB+NB+HI} & \texttt{DM0} & 4.8 & 0.8 & 5.6  \\
			\texttt{IC+BB+NB+HI+CO} & \texttt{DM0} & 4.6 & 1.3 & 5.9  \\
			\hline
			\hline
		\end{tabular}
		\label{tab:AIC_results}
	\end{table}

	\subsection{Choice of Baseline Model}\label{sec:baseline_model}
	The starting point of comparison is not unique and the path towards including more models can be versatile, even improving upon a worse starting point.
	From previous studies \citep[e.g.,][]{Strong2005_gammaconti,Bouchet2011_diffuseCR,Churazov2011_511}, it is known to what extent positron annihilation contributes to the soft $\gamma$-ray spectrum.
	Below $\approx 250$\,keV, the diffuse Galactic continuum emission as well as (un)resolved point sources dominate the spectrum.
	Above the 511\,keV line, direct annihilation `in flight' is expected, but has never been detected \citep[e.g.,][]{Beacom2006_511,Sizun2006_511,Churazov2011_511}, so that these energies can be considered nearly-free of positron annihilation.
	In order to assess annihilation features with respect to different spatial morphologies, a baseline model is established upon which additional components are tested.

	First, we evaluated the AIC for each map individually in `off-annihilation'-bins (excluding $245$--$514$\,keV), and then combined them to obtain a solid estimate of which maps are required.
	We find that the disc-dominated maps IC, HI, and CO, in this sequence, are clearly favoured over any other single map.
	Using the IC templates, we find  $\chi^2 = 589861$ ($\nu = 581570$; $\chi^2/\nu = 1.014$) in these bins combined, suggesting an adequate fit to our data.
	In a second step, we consecutively add these three maps together to identify possible better baselines.
	We conclude that `off-annihilation'-bins are sufficiently described by IC only.
	For the model comparisons in positron-annihilation bins, we thus use IC as a baseline model.

	\begin{figure}%[!ht]
		\centering
		\includegraphics[width=0.8\columnwidth,trim=0.0in 0.15in 0.4in 0.7in, clip=True]{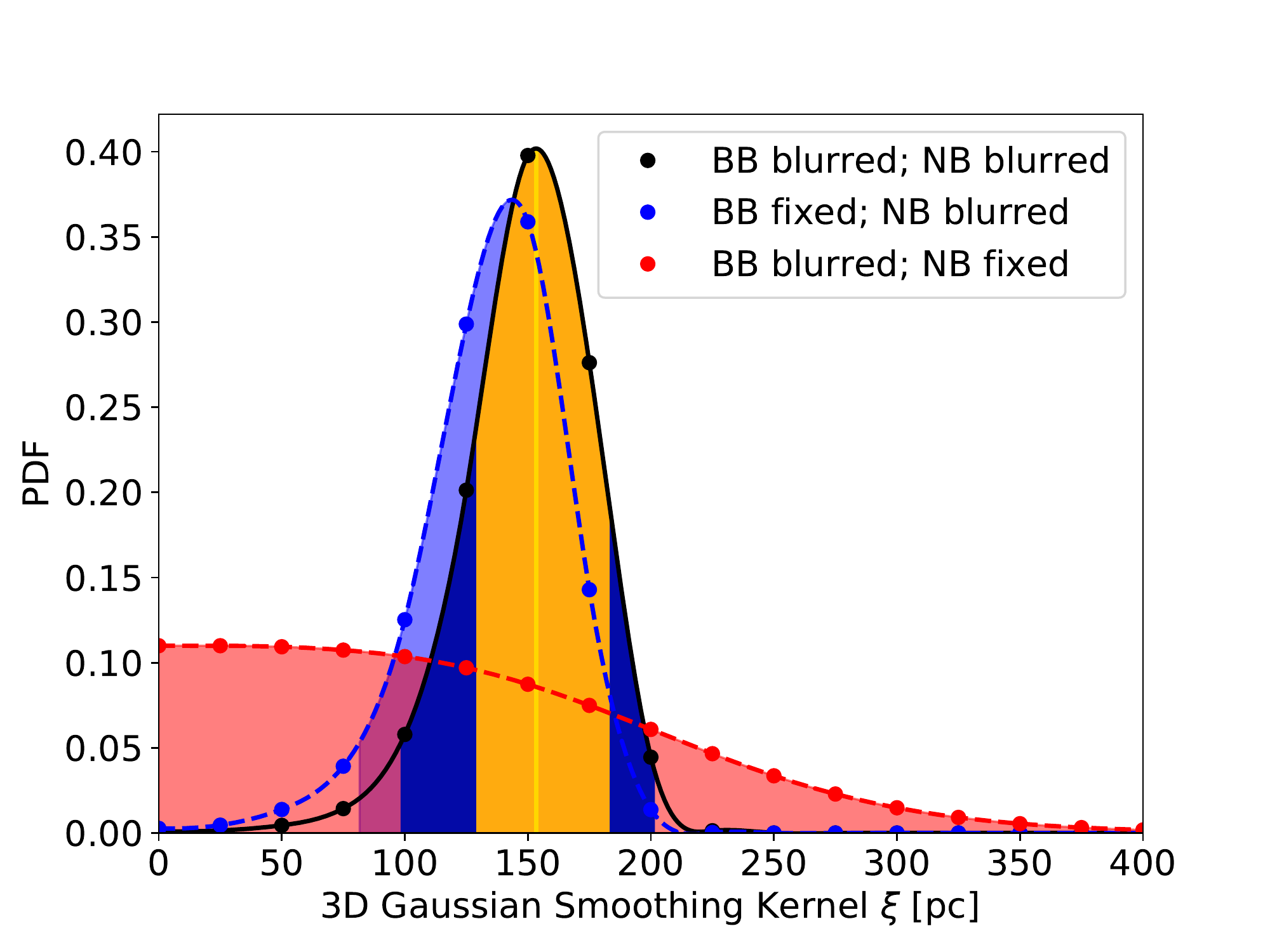}
		\caption{Probability distributions of smoothing widths. \textit{Black}: both templates smoothed by the same $\xi$. \textit{Blue}: only NB blurred. \textit{Red}: only BB blurred. Shown are the $1\sigma$ (orange) and $2\sigma$ (dark blue) intervals for the first case, the $2\sigma$ region for the second (blue), and the $3\sigma$ limit (red) for the third.} 
		\label{fig:blur}
	\end{figure}

	\section{Results}\label{sec:results}
	\subsection{Emission Templates and Combinations}\label{sec:maps_as_they_are}
	Between 245 and 514\,keV, using the stellar template BB+NB improves the fit above the baseline model by $\Delta\mrm{AIC}_{\pm} = 618.2$ (Tab.\,\ref{tab:AIC_results}). 
	This means that the map resembling the projected population of stars in the bulge shows the largest statistical improvement.
	Including the ortho-Ps continuum provides additional power in discriminating different morphologies, as opposed to only using the 511\,keV line.
	While the template DM0 shows the single best improvement when the 511\,keV line is considered in isolation, ortho-Ps provides a much sharper discrimination between individual components that clearly favours the stellar templates.
	The halo template DM2 is significantly worse in all subsequent cases and we only discuss DM0 in the following.
	Adding more maps to IC+BB+NB results in insignificant improvements.
	In particular, a contribution from a DM morphology is not required in the fit.
	Putative annihilation region maps (CO/HI) marginally improve the fit in the annihilation band, and not at all in the 511\,keV line.
	Adding again a DM template when gas maps are included (IC+BB+NB+HI+CO+DM0) results in no improvement.

	\subsection{Similarity between stellar and annihilation morphology}\label{sec:blurring_analysis}
	To determine how similar the distribution of stars (BB+NB) is to the actual annihilation emission, we perform a blurring analysis with these two templates and their combination.
	This increases the number of fitted sky model parameters from three to five.
	First, the same blurring for both templates is used to estimate a common length scale $\xi$.
	We find $\xi = 150 \pm 50$\,pc ($2\sigma$), pointing to a resolvable difference between the initial templates and the annihilation emission (Fig. \,\ref{fig:blur}).
	To assess which component is driving the blurring, we test the two extremal cases for which only one component is blurred, and the other remains unmodified.
	The NB component is responsible for driving the fit because the smoothing kernel for the BB is only constrained to $\xi_{\rm BB} < 420$\,pc ($3\sigma$), whereas from the NB we measure a finite smearing of $\xi_{\rm NB} = 150^{+40}_{-70}$\,pc ($2\sigma$).

	\begin{figure}
		\centering
		\includegraphics[width=0.75\columnwidth,trim=0.20in 0.55in 3.7in 1.2in,clip=true]{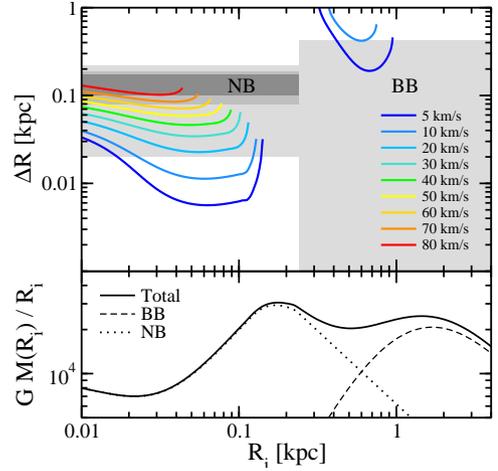}%
		\caption{Displacements from an initial radius $R_i$ in the Galactic bulge (NB \textit{top left}, BB \textit{top right}) potential (\textit{bottom}) for different supernova kick velocities. The gray bands are the regions allowed by our blurring analysis (Dark $1\sigma$, medium $2\sigma$, light $3\sigma$ uncertainties).}%
		\label{fig:kicks}%
	\end{figure}

	\section{Discussion}\label{sec:discussion}
	\subsection{Supernova Kicks}\label{sec:pulsar_kicks}
	A blurred distribution of annihilating positrons could reflect kinematic heating of the initial astrophysical source population due to kicks imparted by asymmetric supernova explosions.
	Thus, the morphology of the population of compact objects that produce positrons might differ from the parent stellar distribution.
	We estimate the average kick velocity $v_k$ that would lead to an extended spatial distribution with radius $R_k$ assuming initially virialised objects at radius $R_i$ with velocity $v_i$ in the gravitational potential of NB and BB,
	\begin{equation}
		\frac{1}{2}v_i^2 = \frac{1}{2} G \frac{M(R_i)}{R_i} \stackrel{\mrm{kick}}{\Longrightarrow} \frac{1}{2}(v_i^2 + v_k^2) = \frac{1}{2} G \frac{M(R_f)}{R_f} \mrm{,}
		\label{eq:virial_init}
	\end{equation}
	\noindent where $M(R)$ is the cumulative mass profile of the bulge up to radius $R$, and $G$ is the gravitational constant.
	We relate the displacement $\Delta R = R_f - R_i$ to the blurring length scale $\xi$, and illustrate $\Delta R$ for different kick velocities in Fig.\,\ref{fig:kicks}.
	Within our $3\sigma$ limits the BB only permits small kick velocities $\lesssim 10\,\mrm{km\,s^{-1}}$.
	In contrast, the NB requires higher kick velocities, $\gtrsim 20\,\mrm{km\,s^{-1}}$ reaching also beyond $\approx 100\,\mrm{km\,s^{-1}}$. 
	Kick velocities around $5$--$15\,\mrm{km\,s^{-1}}$ would match the measurements in both NB and BB ($3\sigma$), but with further restrictions on where the sources would be located.
	In particular this would mean that sources inhabiting only a fraction of either bulge component could contribute positrons supplying the observed signals.

	Sources may be produced in the NB and BB with small kicks.
	Globular clusters, for example, retain populations of millisecond pulsars \citep{Song2021_GC_IC} which must be born with small kicks, $\lesssim$ few tens of $\mrm{km\,s^{-1}}$, given clusters' typical escape velocities.
	An alternative millisecond pulsar formation channel 
	that naturally produces low kick velocities \citep[$\lesssim 30\,\mrm{km\,s^{-1}}$,][]{Tauris2013_pulsarkicksMSP}, is
	the accretion induced collapse of ONe white dwarves.
	Such a population has recently been suggested in the Galactic bulge \citep{Gautam2021_MSPs_GCE}.

	\begin{figure}
		\centering
		\includegraphics[width=0.8\columnwidth,trim=0.0in 0.25in 1.0in 0.9in, clip=true]{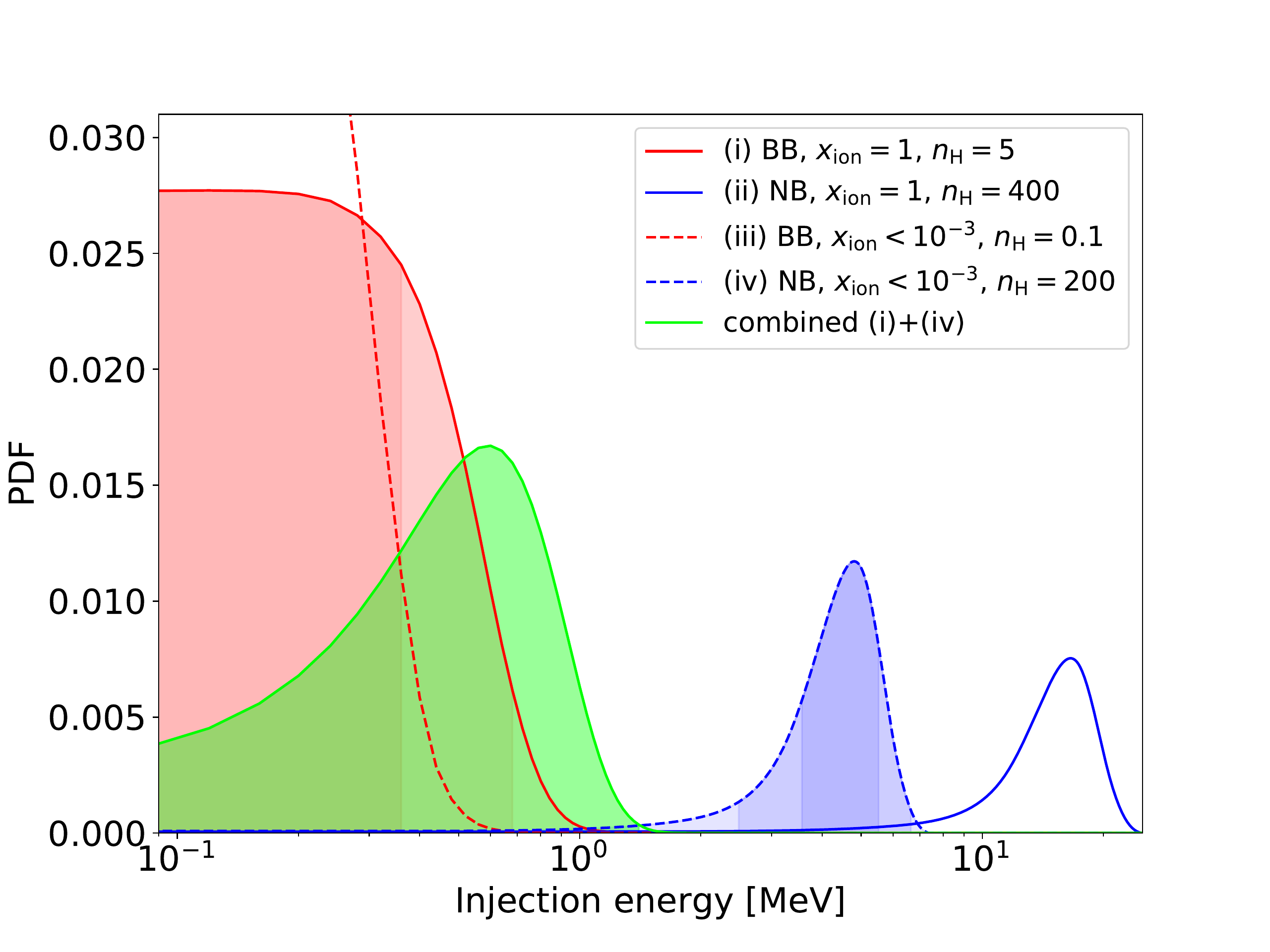}%
		\caption{Allowed parameter space for the NB (blue) and BB (red) as a function of injection energy, particle density, and ionisation fraction. Shown are the extremes (ii) and (iii) with the least and largest overlap (i)+(iv) in green, allowing energies up to $\approx 1.4$\,MeV. (Bands: $1$, $2$, $3\sigma$ bounds).}%
		\label{fig:limits_Einj}%
	\end{figure}

	\subsection{Positron Propagation}\label{sec:positron_propagation}
	Another possibility to obtain a blurred annihilation signal is to have positrons propagating in the ISM before annihilation.
	To test this, we run Monte Carlo simulations as a function of the ISM density $n_{\rm H}$, ionisation fraction $x_{\rm ion}$, and injection energy $E_{\rm inj}$.
	From the canonical ISM phases \citep{Ferriere2001_ISM}, cold neutral (CNM), warm neutral (WNM), warm ionised (WIM), and hot ionised medium (HIM), we determine bounds on the positron injection energy in NB and BB independently.
	In particular, we simulate the interactions of a positron starting at kinetic energy $E_{\rm inj}$ until the  distance travelled between two successive interactions drops below 0.1\,pc.
	We include the cross sections with free and bound electrons, charge exchange, and radiative recombination, energy losses due to ionisation or excitation of hydrogen, plasma losses, bremsstrahlung, synchrotron radiation and IC scattering.
	We use a generic magnetic field strength of $50\,\mrm{\mu G}$ \citep{Ferriere2009_ISM_Bfields,Crocker2010_GalCen_50muGauss}.
	Following the arguments presented by \citet{Jean2009_511ISM}, we assume that various damping processes ensure there is insignificant magnetic structure on the $7 \times 10^7 \ (E_{e^+}/{\rm MeV})(50\,\mu{\rm G}/B)$\,cm gyroradius scale.
	On the basis of \citet{Panther2018_nuclear_outflow} we also discount advective transport in a large-scale wind.
	To account for the uncertain magnetic structure in the bulge, we either (a) follow strictly the results of \citet{Jean2009_511ISM} with inefficient pitch angle scattering, or (b) apply a complete random walk behaviour after each interaction.
	This estimates the maximum (a) and minimum (b) distance spread of positrons, and in return the minimum (a) and maximum (b) injection energy.

	We simulate 1000 Monte-Carlo packets each for logarithmic grids of $n_{\rm H}$ in $0.1$--$10^4\,\mrm{cm^{-3}}$, $x_{\rm ion}$ in $0$--$1$, and $E_{\rm inj}$ in $0.1$--$100$\,MeV for a total of 616 grid points.
	Each distribution then shows final pseudo-random-walk distances of positrons, scattered around 0, with a width that we relate to the blurring length scale $\xi$.
	In Fig.\,\ref{fig:limits_Einj}, we show the limiting injection energies from our measured length scales in NB and BB, assuming a complete random walk.

	For densities in the NB of $200$--$400\,\mrm{cm^{-3}}$ \citep{Launhardt2002_NB}, we find injection energies of $0.9$--$25\,\mrm{MeV}$ reproduce the observed smearing.
	For the BB we consider densities of $0.1$--$5\,\mrm{cm^{-3}}$.
	To explain the maximum blurring scale in the BB of $\lesssim 420\,\mrm{pc}$, positron injection energies of, at most, 1\,MeV are possible for a fully ionised medium at rather high density ($5\,\mrm{cm^{-3}}$), which appears unphysical.
	The resultant annihilation line spectrum in such a medium would contradict the observed spectrum \citep{Guessoum2005_511,Siegert2019_lv511}.
	At lower ionisation fractions, as suggested from spectral measurements \citep{Jean2006_511,Churazov2005_511}, the injection energies would be constrained to values $\lesssim 0.9\,\mrm{MeV}$.
	While this is consistent with the injection energies found from the NB, the large non-zero propagation lengths in the NB would be in tension with the data from the BB.
	If \textit{the same sources} are responsible for positron production in NB and BB, these sources would show injection energies of $\lesssim 1.4$\,MeV.
	The results when using the canonical ISM phases for NB and BB are similar in shape but limit the maximum possible injection energy to $\lesssim 0.4$\,MeV (see appendix\,\ref{sec:ISM_phases_results}). 

	\section{Summary and Conclusion}\label{sec:summary}
	In earlier works \citep[e.g.,][]{Knoedlseder2005_511,Bouchet2011_diffuseCR} it was noted that the 511\,keV emission distribution is reminiscent of the old stellar population.
	Here we have  secured the hypothesised 
	association between annihilating
	positrons and old stars with a broader, quantitative analysis.
	Our spatial model fitting demonstrates that stellar templates for the NB and BB best match the positron annihilation emission as traced by the 511\,keV line and ortho-Ps continuum.
	
	Still, this match is not perfect and can be systematically improved by introducing a characteristic $150 \pm 50$\,pc ($2\sigma$)
	radial, 3D-blurring of the signal with respect to the underlying stars.
	While such a blurring might indicate a systematic displacement of either positron sources (presumably via natal kicks) or ISM transport of positrons, the former would require differently-kicked positron sources in NB and BB.
	In contrast, the transport scenario can accommodate the different degree of apparent blurring in NB and BB given the different ISM conditions they present, albeit with some tuning of parameters.
	Thus, if we hypothesise that the annihilation emission in NB and BB originate from the \textit{same source type}, the ISM propagation scenario is favoured.
	This casts some doubt on in-situ annihilation scenarios \citep[e.g.,][]{Bisnovatyi-Kogan2017_511}.
	
	We determine the distance spread of propagating positrons with initial energy $E_{\rm inj}$ until annihilation when experiencing different ISM conditions.
	In the NB we set constraints on the injection energy of $0.9 \lesssim E_{\rm inj}/\mrm{MeV} \lesssim 25$ ($3\sigma$).
	For the BB we find $E_{\rm inj} \lesssim 1.0\,\mrm{MeV}$ ($3\sigma$), marginally consistent with the NB.
	Again assuming that \textit{the same populations of sources} are responsible for the positrons that annihilate in both the NB and BB, the average injection energy must be at most $1.4$\,MeV, an energy scale that clearly points to a nucleosynthesis origin of the positrons.
	This constraint is in line with previous limits on the injection energy from spectral modelling of the annihilation-in-flight continuum \citep{Beacom2006_511,Sizun2006_511}, suggesting $E_{\rm inj} \lesssim 3$--$7$\,MeV.
	However, the allowed higher energies in the NB would include also relativistic mechanisms, such as pair production from compact objects.
	Finally, higher injection energies in the NB and the resulting larger propagation distances inevitably would lead to outflows into the BB.

	While in-situ annihilation at kicked sources is disfavoured when assuming the same positron producers in NB and BB, scenarios with two different source populations should be explored.
	Dedicated MeV observations and modelling of globular clusters should be considered because at least some compact sources must be born with low kicks.
	Globular clusters are mostly free of interstellar gas so that cooling of positrons is dominated by IC and synchrotron losses.
	A detection of a 511\,keV signal from globular clusters may require the relaxation of our assumption that all positrons in the Galactic bulge originate from the same source population.
	Individual globular clusters would be expected to exhibit 511\,keV line fluxes on the order of $10^{-6}\,\mrm{ph\,cm^{-2}\,s^{-1}}$ if a correlation with GeV emission \citep{Bartels2018_binaries511} could be consolidated.

	\section*{Acknowledgements}
	Thomas Siegert is supported by the German Research Foundation (DFG-Forschungsstipendium SI 2502/3-1).
	RMC acknowledges 
	support from the Australian Government through the Australian Research Council, award
	DP190101258 (shared with Prof. Mark Krumholz).
	Oscar Macias was supported by World Premier International Research Center Initiative (WPI Initiative), MEXT, Japan and by JSPS KAKENHI Grant Numbers JP17H04836, JP18H04340, JP18H04578, and JP20K14463.
	FHP is supported by the Australian Research  Council  (ARC)  Centre  of  Excellence for  Gravitational  Wave  Discovery (OzGrav)  under grant CE170100004.
	Francesca Calore acknowledges 
	support from the "Agence Nationale de la Recherche”, grant n. ANR-19-CE31-0005-01 (PI: F. Calore).
	The work of Shunsaku Horiuchi and Deheng Song is supported by the US Department of Energy Office of Science under award number DE-SC0020262.
	Shunsaku Horiuchi is supported by NSF Grants No.~AST-1908960 and No.~PHY-1914409. This work was supported by World Premier International Research Center Initiative (WPI), MEXT, Japan.

	%%%%%%%%%%%%%%%%%%%%%%%%%%%%%%%%%%%%%%%%%%%%%%%%%%
	\section*{Data Availability}
	The data underlying this article will be shared on reasonable request to the corresponding author.
	
	%The inclusion of a Data Availability Statement is a requirement for articles published in MNRAS. Data Availability Statements provide a standardised format for readers to understand the availability of data underlying the research results described in the article. The statement may refer to original data generated in the course of the study or to third-party data analysed in the article. The statement should describe and provide means of access, where possible, by linking to the data or providing the required accession numbers for the relevant databases or DOIs.

	
	
	\bibliographystyle{mnras}
	\bibliography{thomas} % if your bibtex file is called example.bib

	%\begin{thebibliography}{99}
	%\end{thebibliography}
	
	%%%%%%%%%%%%%%%%%%%%%%%%%%%%%%%%%%%%%%%%%%%%%%%%%%
	
	%%%%%%%%%%%%%%%%% APPENDICES %%%%%%%%%%%%%%%%%%%%%
	
	\appendix

	\section{Morphological Emission Templates}\label{sec:emission_templates}
	
\begin{figure*}
		\centering
		\begin{subfigure}
			\centering
			\includegraphics[width=0.33\textwidth,trim=0.44in 2.02in 0.12in 2.34in,clip=true]{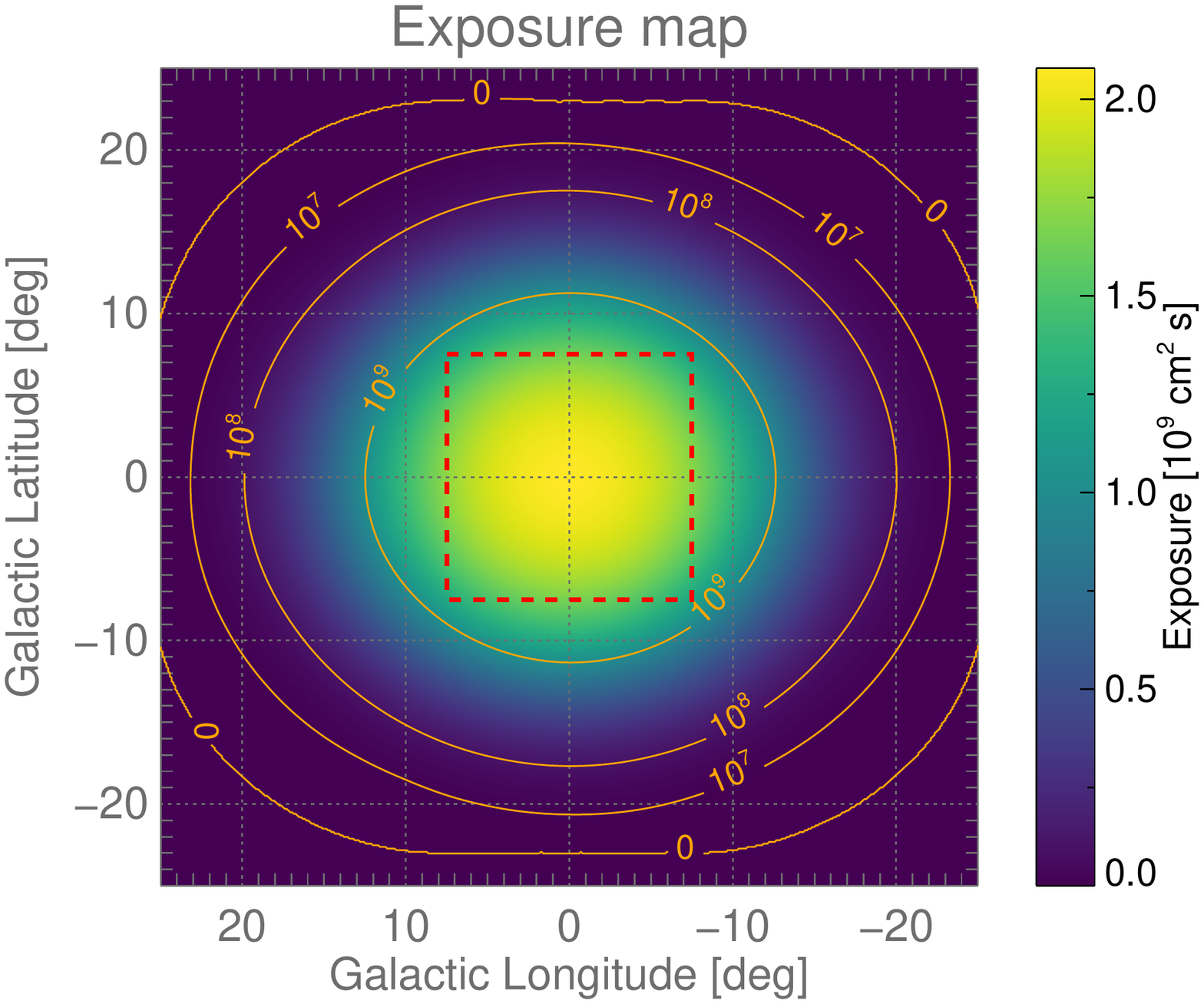}
			\includegraphics[width=0.33\textwidth,trim=0.44in 2.03in 0.12in 2.34in,clip=true]{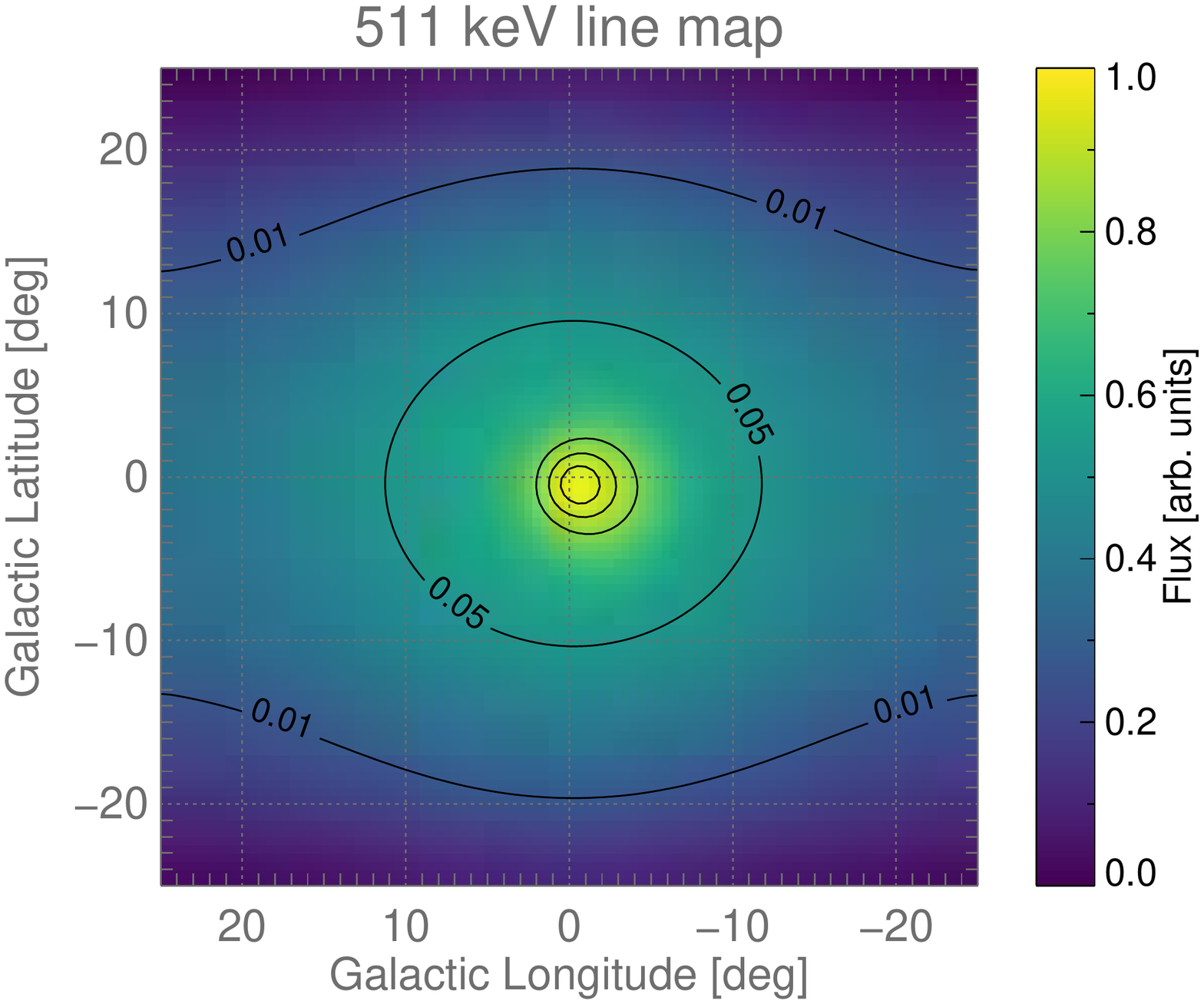}
			\caption{Exposure map (left) and empirical 511\,keV map from \citet{Siegert2016_511} (log, right)}\label{fig:expo_map}\label{fig:511_map}
		\end{subfigure}
		\begin{subfigure}
			\centering
			\includegraphics[width=0.25\textwidth,trim=0.44in 2.03in 1.29in 2.34in,clip=true]{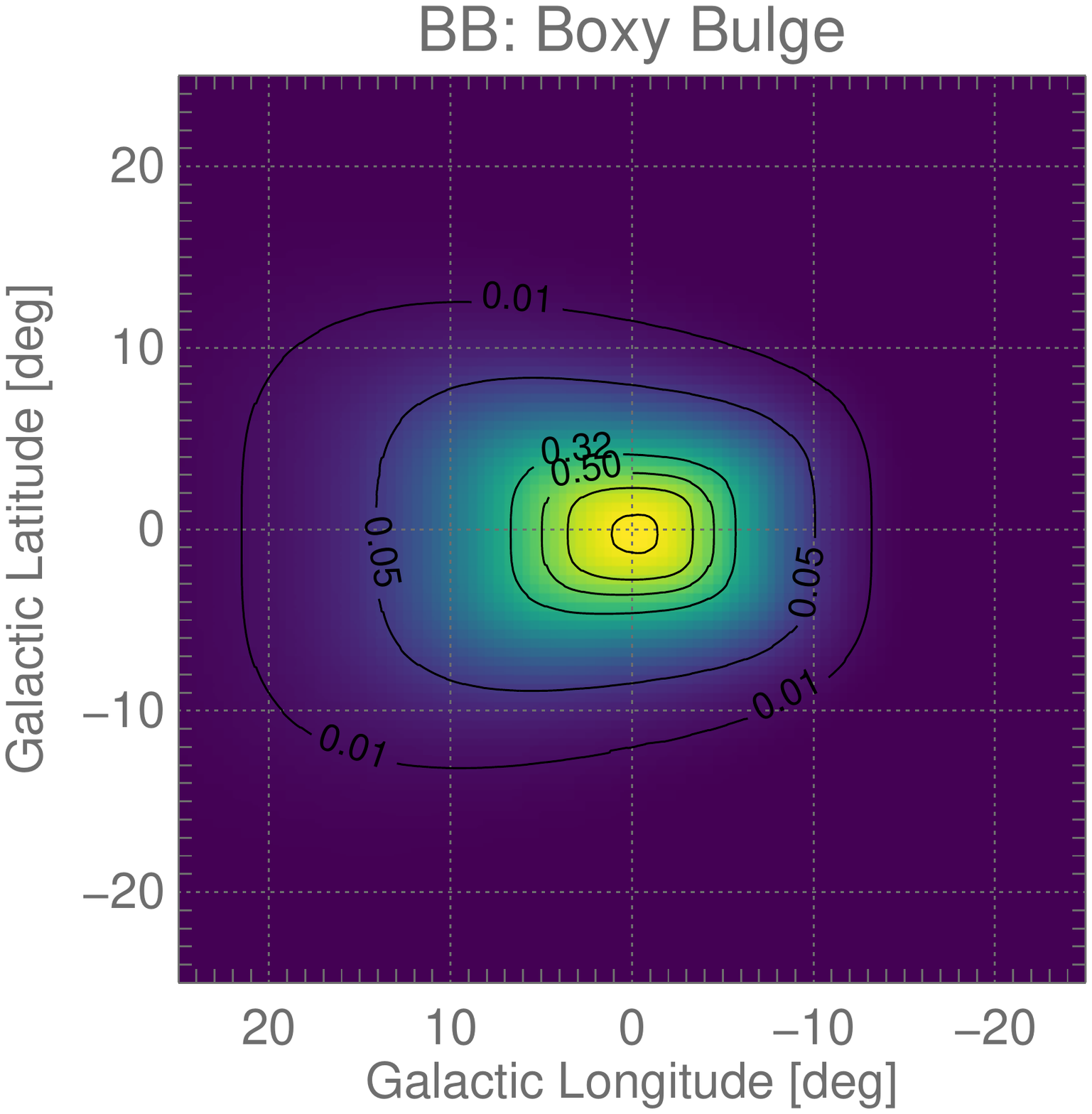}
			\includegraphics[width=0.25\textwidth,trim=0.44in 2.03in 1.29in 2.34in,clip=true]{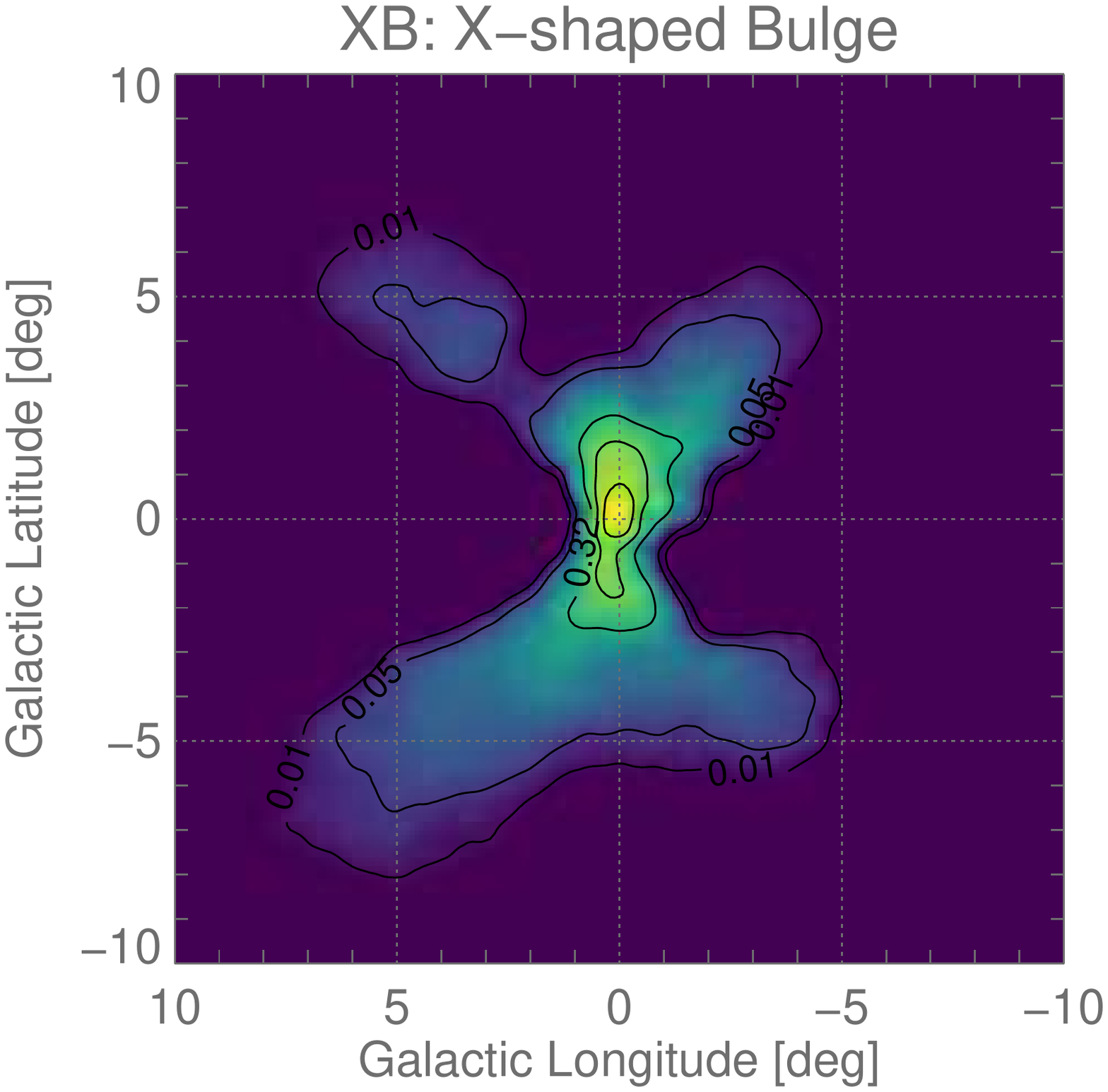}
			\includegraphics[width=0.25\textwidth,trim=0.44in 2.03in 1.29in 2.34in,clip=true]{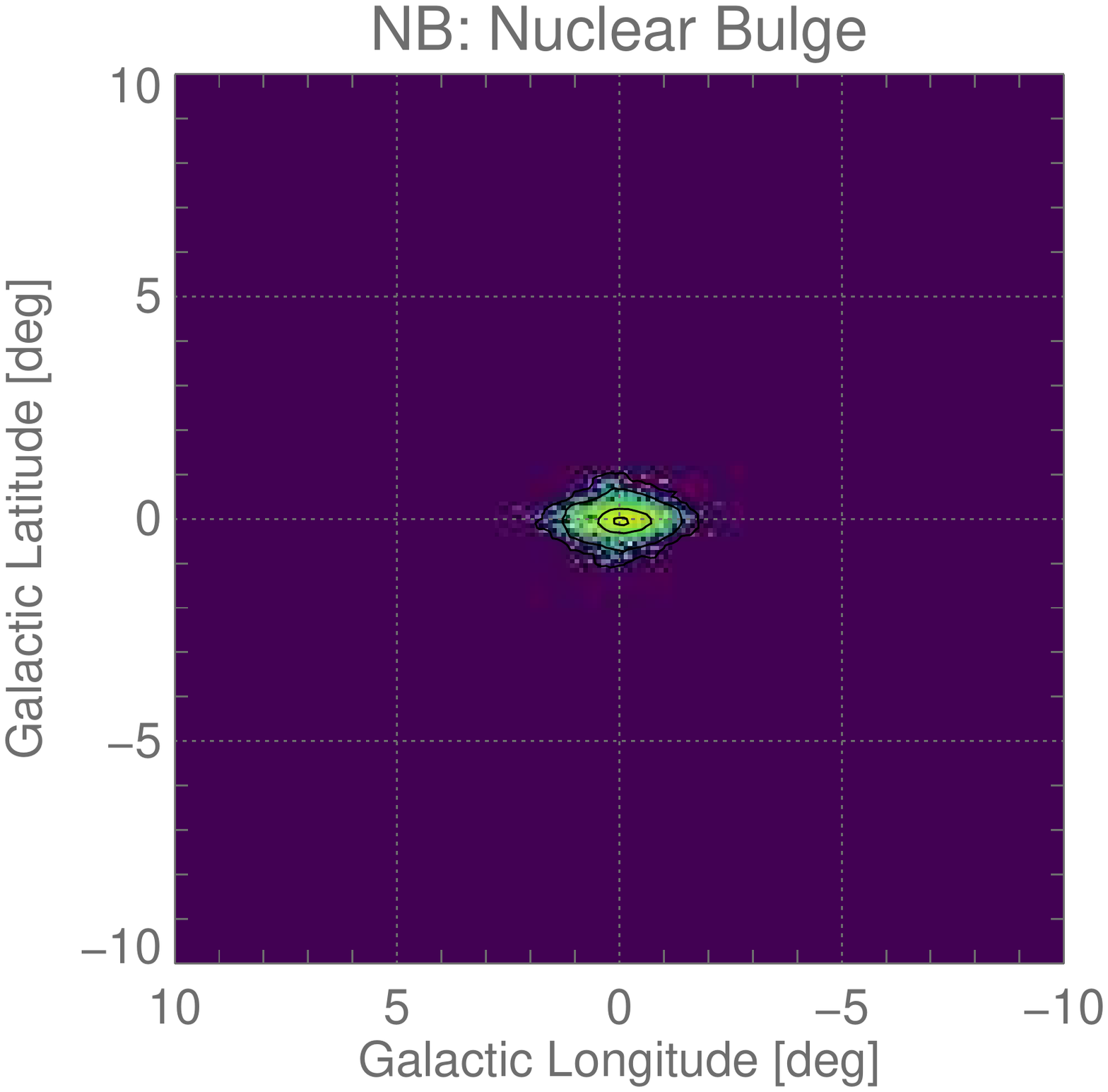}
			\caption{Bulge maps: Boxy Bulge (asinh, left), X-bulge (asinh, zoom, middle), and Nuclear Bulge (log, zoom, right).}\label{fig:BB_map}\label{fig:XB_map}\label{fig:NB_map}
		\end{subfigure}
		\begin{subfigure}
			\centering
			\includegraphics[width=0.25\textwidth,trim=0.44in 2.03in 1.29in 2.34in,clip=true]{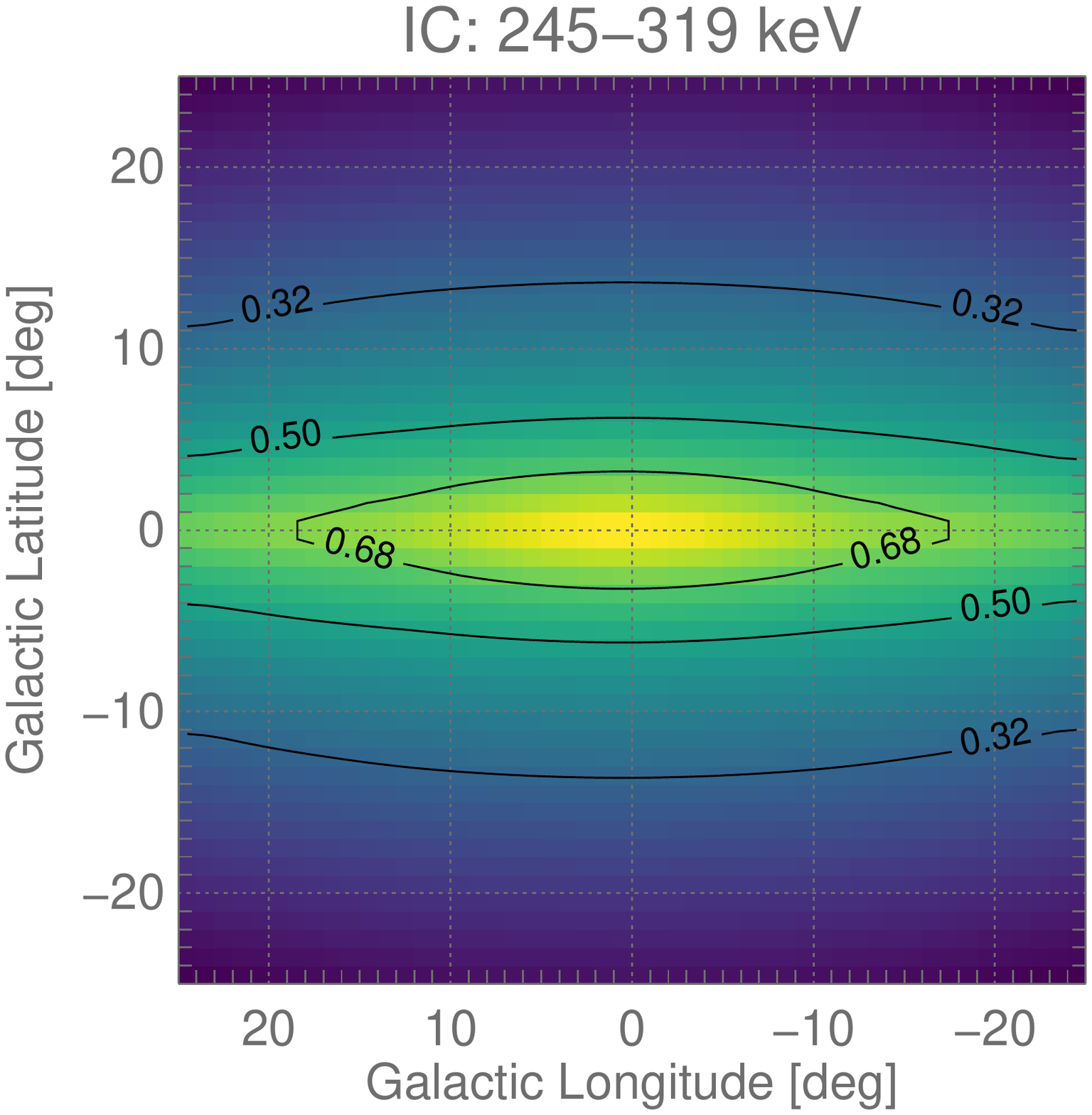}
			\includegraphics[width=0.25\textwidth,trim=0.44in 2.03in 1.29in 2.34in,clip=true]{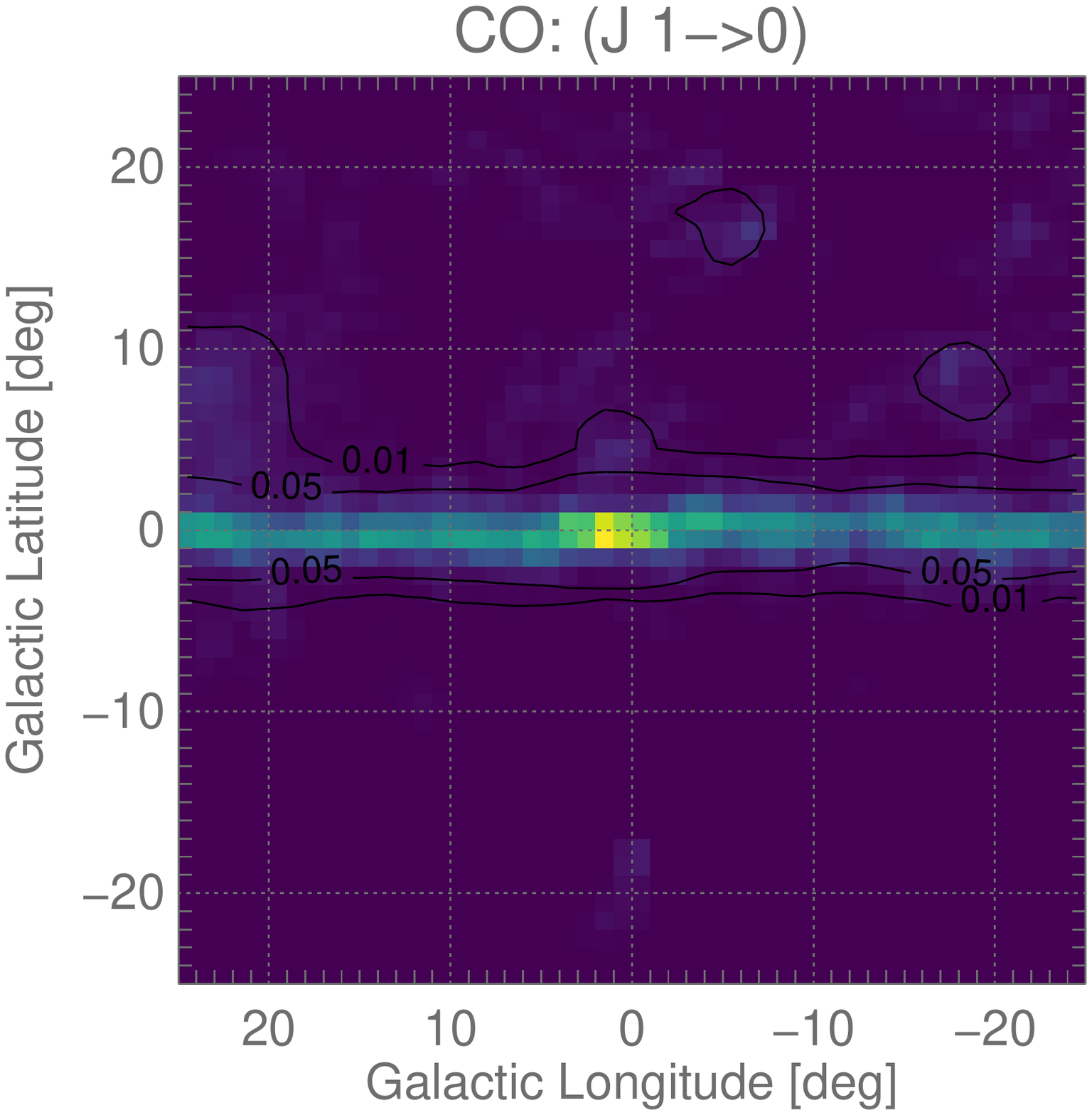}
			\includegraphics[width=0.25\textwidth,trim=0.44in 2.03in 1.29in 2.34in,clip=true]{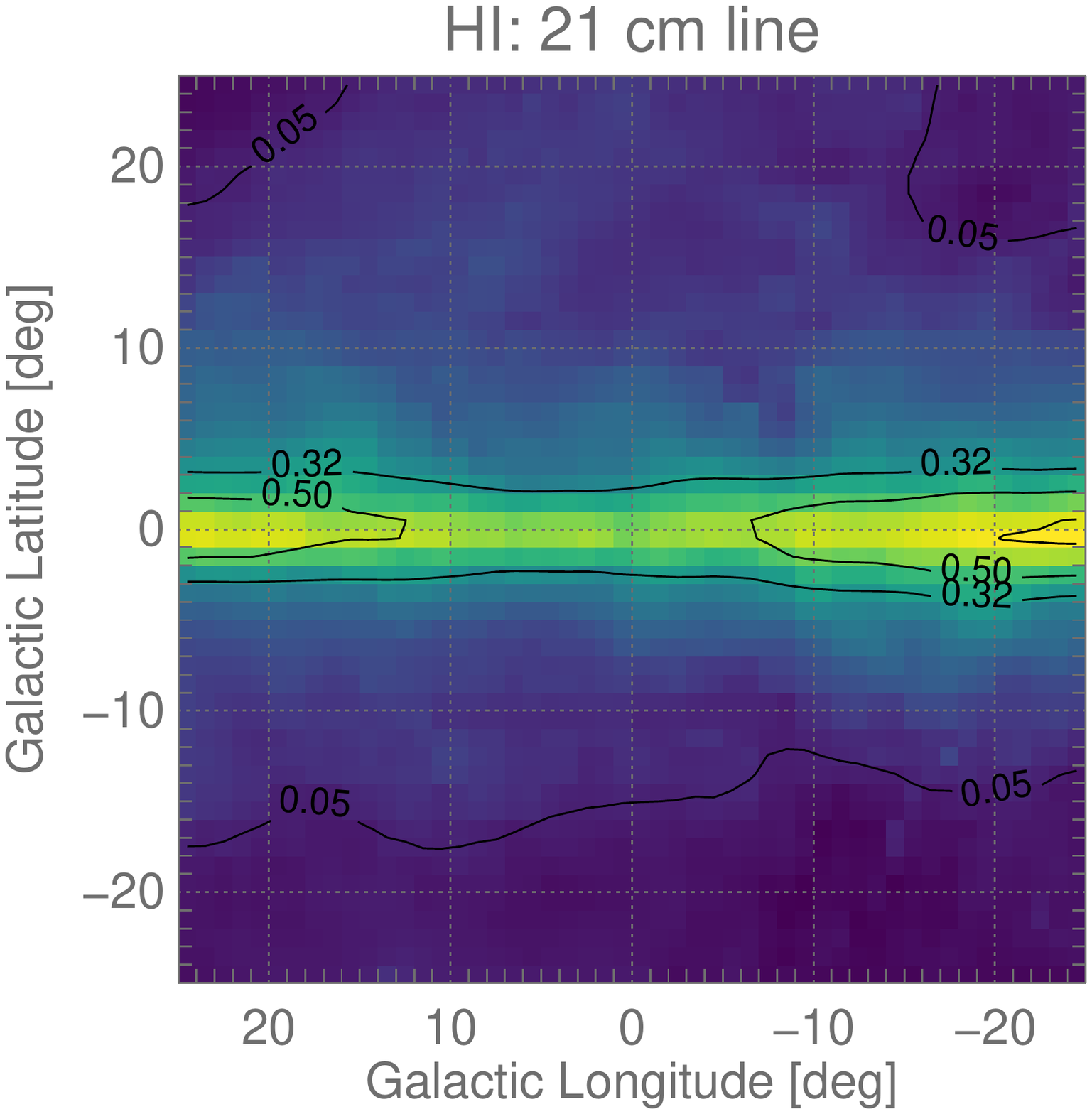}
			\caption{Disk maps: Inverse Compton (asinh, left), Planck CO (asinh, middle), and HI 21\,cm survey (asinh, right).}\label{fig:IC_map}\label{fig:CO_map}\label{fig:HI_map}
		\end{subfigure}
		\begin{subfigure}
			\centering
			\includegraphics[width=0.25\textwidth,trim=0.44in 2.03in 1.29in 2.34in,clip=true]{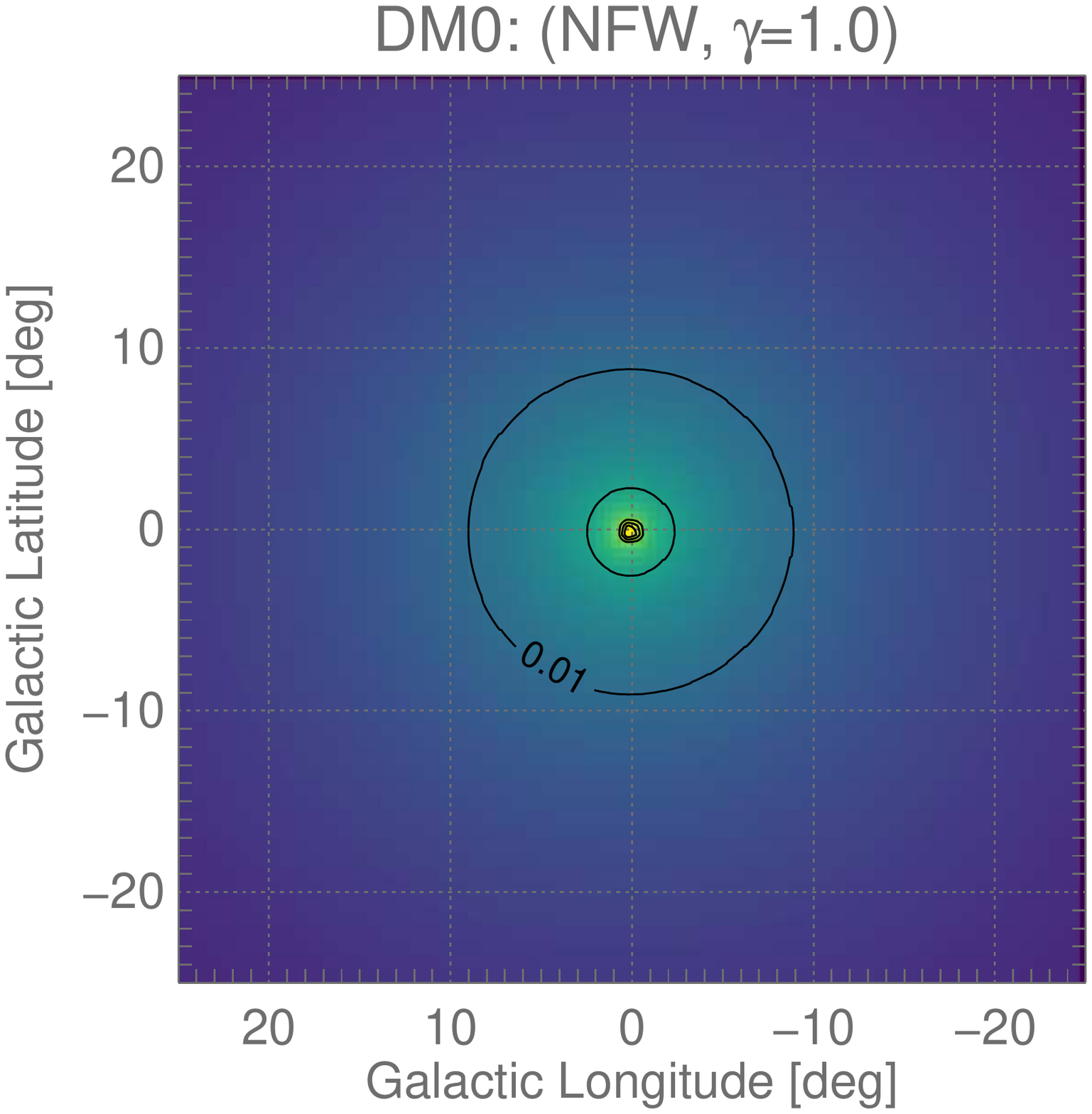}
			\includegraphics[width=0.25\textwidth,trim=0.44in 2.03in 1.29in 2.34in,clip=true]{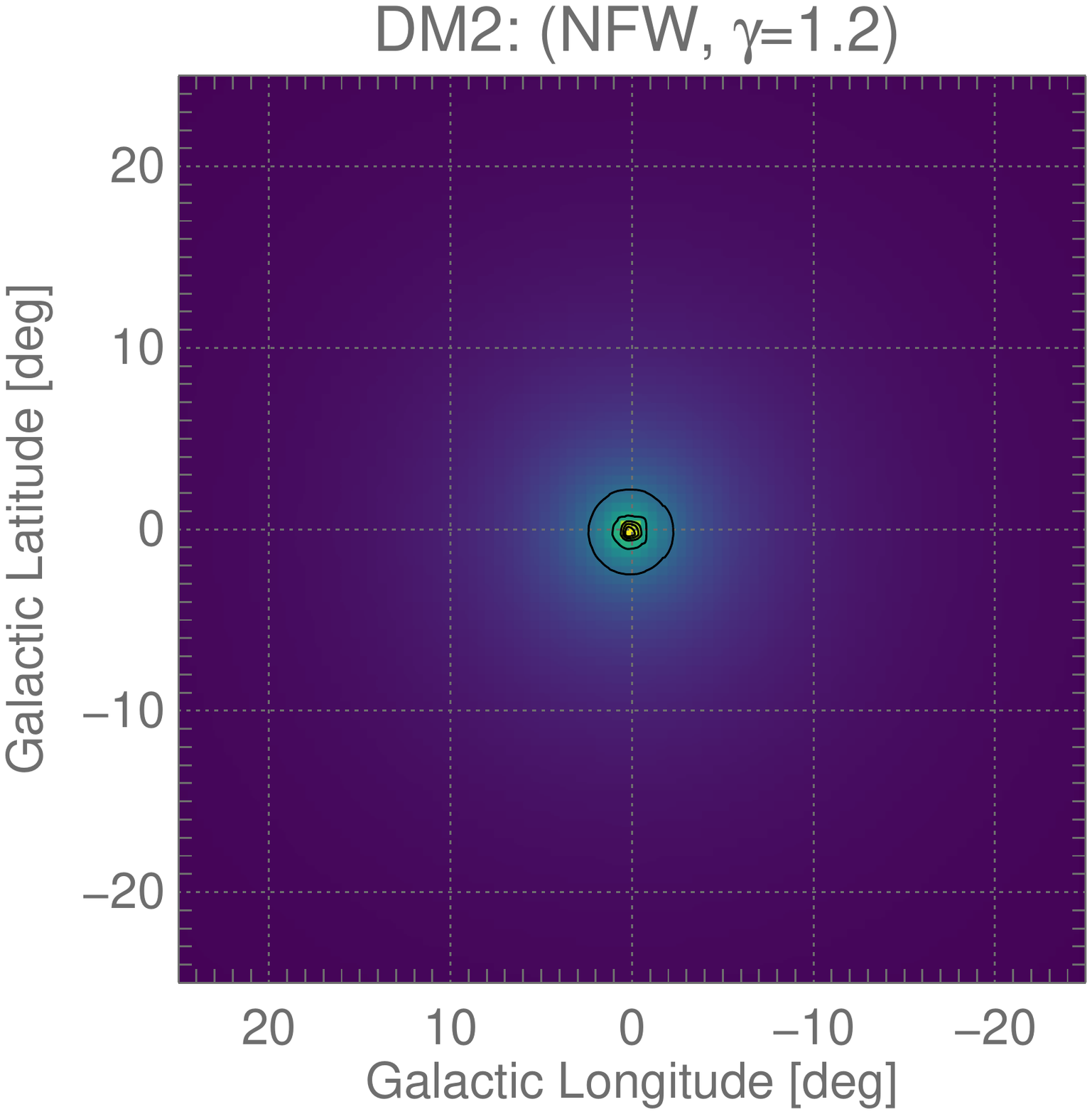}
			\includegraphics[width=0.25\textwidth,trim=0.44in 2.03in 1.29in 2.34in,clip=true]{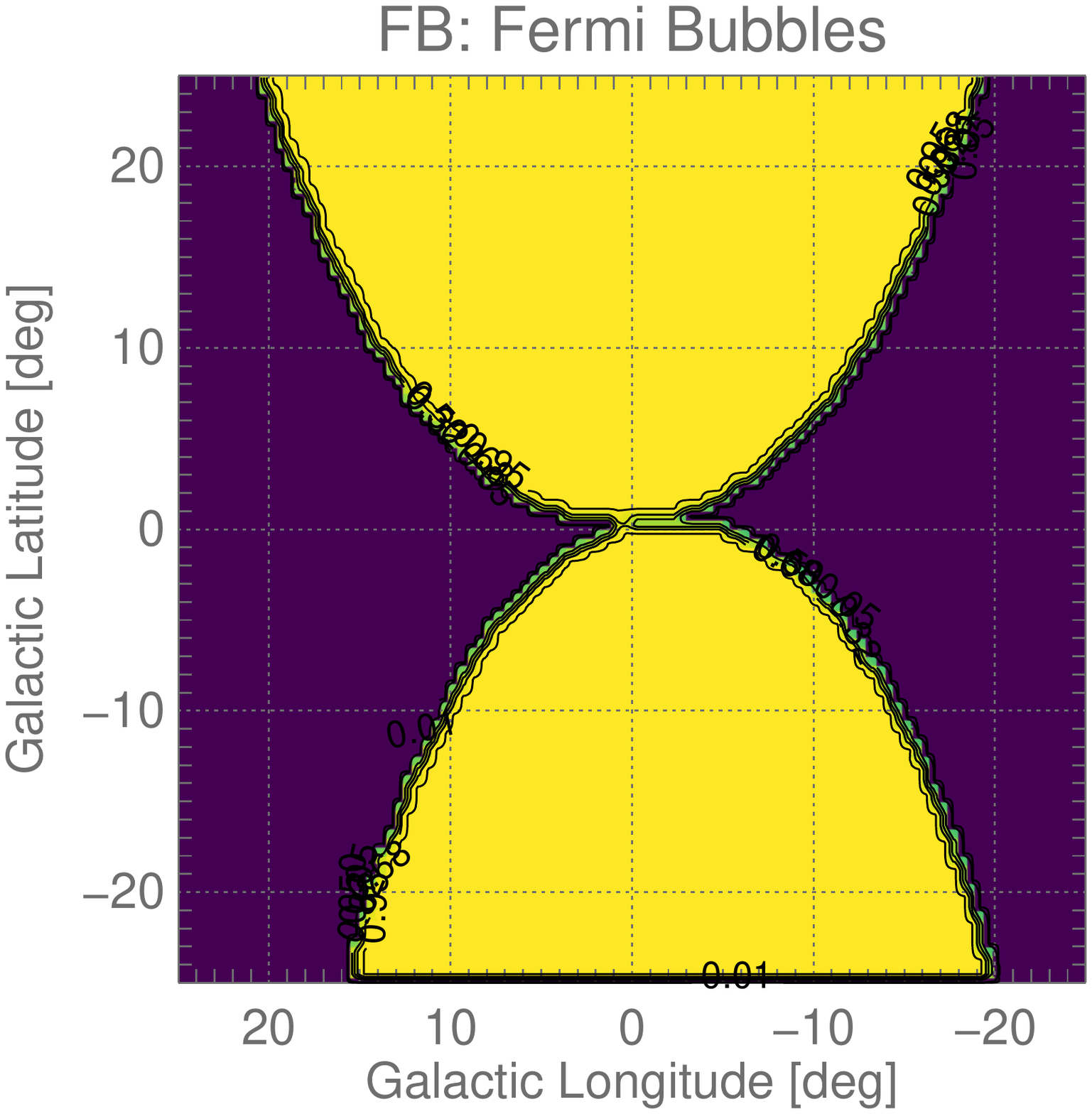}
			\caption{Halo maps: NFW with $\gamma=1.0$ (log, left), NFW with $\gamma=1.2$ (log, middle), and Fermi Bubbles (log, right).}\label{fig:DM0_map}\label{fig:DM2_map}\label{fig:FB_map}
		\end{subfigure}
		\caption{Exposure map (top left) and template maps. Observation pointings were chosen inside the red rectangle ($15^{\circ} \times 15^{\circ}$), resulting in exposure beyond $20^{\circ}$ from the centre. The maps are scaled from 0 to 1 (see top right). The relative scaling is noted in brackets.}
	\end{figure*}

	In total, we test nine individual template maps and combinations thereof (cf. Tab.\,1).
	The individual maps are illustrated in Figs.\,\ref{fig:511_map} to \ref{fig:FB_map}.
	These templates may trace either populations associated with emission in such wavelength bands or the emitting population itself.
	For example, a halo morphology can test for an old population of objects ($\gtrsim 1\,\mrm{Gyr}$), such as type Ia supernovae which produce positrons through radioactively decaying nucleosynthesis ejecta, or the stars of the halo themselves which show flares and positron-annihilation signatures directly.
	
	We test for bulge populations using a Boxy Bulge (\texttt{BB}) template, an X-shaped Bulge (\texttt{XB}), and the Nuclear Stellar Bulge (\texttt{NB}) population as was used earlier \cite[e.g., ][]{Macias2018_LATGeV,Bartels2018_GeVexcess_stars} for Fermi/LAT GeV analysis in the Galactic bulge region.
	Disc contributions are tested via the Planck \texttt{CO} ($J=1 \rightarrow 0$) map \citep{Planck2016_foregrounds}, a line-of-sight-integrated \texttt{HI} map from 21\,cm radio observations \citep{Dickey1990_HI}, and energy-dependent Inverse Compton (\texttt{IC}) scattering template maps from GALPROP calculations \citep{Strong2007_GALPROP}.
	The halo is described either as a Navarro-Frenk-White \citep{Navarro1997_NFW} dark matter halo profile with a $\gamma$-parameter of $1.0$ (\texttt{DM0}) or $\gamma=1.2$ (\texttt{DM2}), or an isotropic model template for emission from the Fermi Bubbles \citep[\texttt{FB}][]{Su2010_fermibubbles}.
	
	In cases for which the emissivity can be modelled through a 3D-density distribution, we perform line-of-sight integrations to determine the flux per steradian, $F(l,b)$, on a $(50^{\circ} \times 50^{\circ})$-sized grid of pixels with solid angles $(0.5^{\circ} \times 0.5^{\circ})$: 
	
	\begin{equation}
		F(l,b) = \frac{1}{4\pi} \int_{los} \rho(x(s),y(s),z(s)) ds\mrm{.}
		\label{eq:los_integral}
	\end{equation}
	
	\noindent Here, $s(l,b)$ is the line of sight (los) along galactic coordinates $(l,b)$.
	For any such map that is parametrised in 3D, we also calculate the luminosity from a measured flux value by 
	
	\begin{equation}
		L = \int_{d\Omega} \int_{los} \rho(x(s),y(s),z(s)) s^2 ds d\Omega\mrm{.}
		\label{eq:luminosity}
	\end{equation}
	
	\noindent This means, we do not have to rely on approximate or effective distances to a diffuse map as is the case for 2D templates.

	\subsection{Boxy Bulge}
	
	The \texttt{BB} was derived by \citet{Freudenreich1998_BoxyBulge_COBE}.
	Triaxial models of the form
	
	\begin{eqnarray}
		\rho_{BB}(x,y,z)/\rho_{0,\rm BB} = & \\\nonumber \left \{
		\begin{aligned}
			&\sech^2\left(R_s\right), && \text{for}\ R \leq R_{\rm end} \\
			&\sech^2\left(R_s\right) \times \exp\left(-(R_s-R_{\rm end})^2/h_{\rm end}^2\right), && \text{for}\ R > R_{\rm end} \\
		\end{aligned} \right. &
		\label{eq:BB_formula}
	\end{eqnarray} 
	
	\noindent where $R_s$ describes an ellipsoid,
	
	\begin{equation}
		R_s = \left\{ \left[ \left(\frac{|x|}{a_x}\right)^{C_{\bot}} + \left(\frac{|y|}{a_y}\right)^{C_{\bot}} \right]^{\frac{C_{||}}{C_{\bot}}} + \left(\frac{|z|}{a_z}\right)^{C_{||}} \right\}^{\frac{1}{C_{||}}}\mrm{,}
	\end{equation}
	
	\noindent were fitted to COBE/DIRBE near-infrared data in four bands between $1.25$ and $4.9\,\mrm{\mu m}$.
	Here, we use the best-fitting `Model S' to describe the rotated bar of the Milky Way with a tilt angle of $\theta_0 = 14^{\circ}$, an axis-ratio in units of kpc of $a_x : a_y : a_z = 1.70 : 0.64 : 0:44$, and a scale height of 461\,pc.
	The \texttt{BB} thus describes star light and traces the population of K and M (giant) stars in Milky Way bulge.
	This bar model is truncated at a galactocentric radius of $3.5\,\mrm{kpc}$ at the inner edge of the molecular cloud ring of the Milky Way.

	\subsection{X-Bulge}
	
	The X-shaped bulge \citep[\texttt{XB},][]{Ness2016_Xbulge_WISE} has been extracted from WISE W1 and W2 data at $3.4$ and $4.6\,\mrm{\mu m}$, respectively.
	It is argued that the Milky Way bulge is morphologically X-shaped, following from expectations of dynamical models and observations of other barred galaxies.
	Further, the population of stars leading to this shape due of their orbital motion contribute to about 40--45\% of the total bulge mass \citep{Portail2015_MilkyWayorbots_Xbulge}.
	This fraction can be used to estimate the luminosity of the \texttt{XB} template, as it is not parametrised in 3D.
	The arms of the WISE \texttt{XB} (see Fig.\,\ref{fig:XB_map}) appear asymmetric, being longer for positive longitudes ($l \lesssim 7.5^{\circ}$) than for negative longitudes ($l \gtrsim -5^{\circ}$).
	This is interpreted as a projection effect of the bulge being $27^{\circ}$ tilted to the line of sight \citep{Wegg2013_bulgebar}.
	The \texttt{XB} is thus said to trace faint red clump stars with its short arms more distant to the Sun and bright red clump with its long arms closer to the Sun \cite[][cf. discussion about the `double red clump' of the Milky Way]{Lee2018_DoubleRedClump}.
	The authors note that the remaining WISE bands W3 and W4 at $12$ and $22\,\mrm{\mu m}$, respectively, mainly trace dust between $\approx 100$ and $250\,\mrm{K}$ rather than star light and show no enhanced X-shape.

	\subsection{Nuclear Stellar Bulge}
	
	\citet{Nishiyama2013_NuclearBulge_PlasmaMagnet} constructed a stellar number density map of the central $6^{\circ} \times 2^{\circ}$ of the Milky Way in the near-infrared, at $1.63\,\mrm{\mu m}$ ($H$) and $2.14\,\mrm{\mu m}$ ($K_S$).
	In these studies, a colour-magnitude diagram was used to remove foreground sources by cutting the blue $H-K_S$ colours.
	For each star in the field, the interstellar extinction is also taken into account.
	Their limiting magnitude for the stellar number density map is $K_S<10.5$.
	After removing remaining Galactic disc contributions \citep{Macias2018_LATGeV}, this procedure resulted in the template map as shown in Fig.\,\ref{fig:NB_map}, which includes the nuclear stellar disc (NSD) as well as the nuclear stellar cluster (NSC), combined to be known as the nuclear stellar bulge (\texttt{NB}).
	The scale height of the \texttt{NB} is $0.32^{\circ}$, corresponding to 45\,pc at a distance of 8.128\,kpc to the Galactic centre.
	The longitude profile is well described by a broken power-law with indices $-0.3$ up to $|l| \lesssim 0.7^{\circ}$ and $-1.9$ beyond.
	The maximum number of stars per $\mrm{arcmin^2}$ is between 100 and 120.
	The longitudinal profile appears centred at $l=-0.17^{\circ}$ and slightly skewed towards negative longitudes.
	The latitude profile is symmetric around that longitude.
	A convolution with the point-spread function of SPI results in an apparent peak at $l \approx -0.35^{\circ}$.
	We note that the \texttt{NB} in near-infrared traces the distribution of an old stellar population, with ages older than $1\,\mrm{Gyr}$.
	In addition, it is concluded from colour-magnitude computations, that the \texttt{NB} has a different star formation history compared to the Galactic bulge, in a way that the \texttt{NB} has formed stars over its entire lifetime, resulting into more bright stars.
	
	As an alternative to the projected number of stars along the line of sight, and to perform additional analyses with this component (see blurring analysis in Sec.\,\ref{sec:blurring_analysis}), we use the 3D-density distribution by \citet{Launhardt2002_NB}, \citep[see also][]{Bartels2018_GeVexcess_stars}.
	In this work, the NSD is given by
	
	\begin{eqnarray}
		\rho_{\rm NSD}(x,y,z) = & \\\nonumber \left \{
		\begin{aligned}
			&\rho_{0,\rm NSD} \left(\frac{R}{R_{\rm NSD}}\right)^{-0.1} \exp\left(-\frac{|z|}{z_{\rm NSD}}\right), && \text{for}\ R/\mrm{pc} < 120 \\
			&\rho_{1,\rm NSD} \left(\frac{R}{R_{\rm NSD}}\right)^{-3.5} \exp\left(-\frac{|z|}{z_{\rm NSD}}\right), && \text{for}\ 120 \leq R/\mrm{pc} < 220 \\
			&\rho_{2,\rm NSD} \left(\frac{R}{R_{\rm NSD}}\right)^{-10} \exp\left(-\frac{|z|}{z_{\rm NSD}}\right), && \text{for}\ R/\mrm{pc} \geq 220 
		\end{aligned} \right. &
		\label{eq:NSD_formula}
	\end{eqnarray} 
	
	\noindent where $R^2 = x^2 + y^2 + z^2$, $R_{\rm NSD} = 1$\,pc, $z_{\rm NSD} = 45$\,pc, and $\rho_{0,\rm NSD} = 301\,\mrm{M_{\odot}\,pc^{-3}}$.
	The mass of the NSD within 120\,pc is set to $8 \times 10^{8}\,\mrm{M_{\odot}}$ \citep{Launhardt2002_NB}, which then defines $\rho_{1,\rm NSD}$ and $\rho_{2,\rm NSD}$ through continuity conditions. The NSC is described by
	
	\begin{eqnarray}
		\rho_{\rm NSC}(x,y,z) = & \\\nonumber \left \{
		\begin{aligned}
			&\rho_{0,\rm NSC} \left[1 + \left(\frac{R}{R_{\rm NSC}}\right)^2\right]^{-1}, && \text{for}\ R/\mrm{pc} < 6 \\
			&\rho_{1,\rm NSC} \left[1 + \left(\frac{R}{R_{\rm NSC}}\right)^3\right]^{-1}, && \text{for}\ 6 < R/\mrm{pc} \leq 200 \\
			&0, && \text{for}\ R/\mrm{pc} > 200 
		\end{aligned} \right. &
		\label{eq:NSC_formula}
	\end{eqnarray} 
	
	\noindent with $R_{\rm NSC} = 0.22$\,pc and $\rho_{0,\rm NSC} = 3.3 \times 10^{6}\,\mrm{M_{\odot}\,pc^{-3}}$.
	The density $\rho_{1,\rm NSC}$ is given by a continuous profile at 6\,pc.
	
	Comparing the projected stellar number with the line-of-sight-integrated density distribution, we find an absolute log-likelihood difference of $2.3$ in favour of the latter.
	Such an improvement is not significant.
	It however shows the variance of the likelihood fits by the use of different models that intend to describe the same.
	For further discussions, we use the line-of-sight-integrated version as it also provides luminosity information.

	\subsection{Molecular CO Emission}
	
	We use the \texttt{CO} ($J=1 \rightarrow 0$, Fig.\,\ref{fig:CO_map}) emission at $115\,\mrm{GHz}$ to trace molecular gas in the Milky Way, here CO itself and especially $\mrm{H_2}$.
	Molecular hydrogen clouds are expected to be one of the major annihilation sites for positrons, and which would result in very distinct properties of the annihilation spectrum \citep{Guessoum2005_511}.
	It is expected that less Positronium is formed in flight in neutral molecular hydrogen ($90$--$93\,\%$) than in neutral atomic hydrogen ($95$--$98\,\%$), which is a direct result of the larger direct annihilation cross section of $\mrm{H_2}$ and the higher threshold energy to capture an electron from the molecule (8.6\,eV) than from the atom (6.8\,eV).
	As a result, the 511\,keV line width would appear broader for $\mrm{H_2}$ (direct: 1.71\,keV; p-Ps: 6.4\,keV) than for H (direct: 1.56\,keV; p-Ps in flight: 5.8\,keV).
	This comparison applies to a cold medium with $T \lesssim 100\,\mrm{K}$ and Positronium formation in flight.
	In consequence, the \texttt{CO} map would trace the natural positron sinks rather than their sources.
	Here, we use the Planck \texttt{Commander} $J=1 \rightarrow 0$ map which shows the strongest molecular line signal in the Planck 100, 217 and 353\,GHz bands \citep{Planck2016_foregrounds}.
	We note that there are systematic flux differences with respect to the CO survey by \citet{Dame2001_COsurvey}.
	This absolute flux is irrelevant for our analysis, as we fit the intensity to our raw SPI data and use the \texttt{CO} map only as a tracer of where positrons could annihilate.
	
	The \texttt{CO} map is thin in latitude with a scale height of less than $1^{\circ}$, approximately 70--100\,pc.
	Towards the Galactic centre, it also shows close-by emission regions, such as the Scorpius-Centaurus region at higher latitudes, as well as Ophiuchus towards $|l| \lesssim 10^{\circ}$ and Aquila at $l \gtrsim 20^{\circ}$, all at distances of less than 200\,pc.
	The centre appears enhanced, due to many giant molecular clouds along the line of sight.
	However, \texttt{CO} may also show sources of positrons, for example as created by pion-production from cosmic-ray interactions with dense medium \citep{Agaronyan1981_511}.
	If the positrons slow down sufficiently fast for Positronium to be formed, the \texttt{CO} (and $\mrm{H_2}$) molecular clouds could be both production and annihilation sites.
	The still unobserved annihilation in flight continuum of fast (relativistic) positrons with the ISM \citep{Beacom2006_511,Sizun2006_511} would also be expected on the boundaries of such clouds as traced by the \texttt{CO} emission map.

	\subsection{Atomic HI Emission}
	
	At 21\,cm photon wavelength, the \texttt{HI} emission map, Fig.\,\ref{fig:HI_map}, shows the hyperfine transition of neutral hydrogen.
	Also these regions are expected to be prominent positron-annihilation sites as the particle density and relative abundance is very large.
	From previous studies, it has been suggested that most positrons annihilate in warm neutral and warm ionised phases of the ISM with about equal shares \citep{Jean2006_511}.
	Here warm means an electron temperature of $T_e = 8000\,\mrm{K}$ at an electron density of $n_e = 0.1\,\mrm{cm^{-3}}$.
	The warm neutral ISM phase would be directly traced by the \texttt{HI} map, and \citet{Jean2006_511} also estimated an upper limit on annihilation in the cold ISM phase of up to $27\,\%$ which would also be seen at the regions of \texttt{HI} clouds.
	The \texttt{HI} emission morphology, however, is dissimilar to what the 511\,keV line studies revealed during the last 20 years \citep{Purcell1997_511,Knoedlseder2005_511,Siegert2016_511,Siegert2019_lv511,Skinner2014_511,Churazov2011_511}.
	Simulations, on the other side, expect the annihilation emission to be closer to what is seen in \texttt{HI}, for example, also with a more patchy morphology and features such as the spiral arms of the Milky Way \citep{Alexis2014_511ISM,Panther2018_pos_transport}.
	We use the \texttt{HI} emission map to include this paradigm of positrons annihilating in large fractions in neutral phases of the ISM. 
	
	The line-of-sight-integrated \texttt{HI} map shows a larger scale height compared to \texttt{CO} emission \citep{Dickey1990_HI}.
	\citet{Kalberla2009_HI} note that individual \texttt{HI} structures may reach up to several kpc in vertical height, and that the disc emission is made of spurs, chimneys, and shells.
	The latter contributes to about 50\% of the total emission up heights of $\approx 500$\,pc.
	The scale height increases exponentially with distance from the Galactic centre, with about 200--300\,pc between 3 and 10\,kpc, and more than 1\,kpc beyond 25\,kpc.
	The rotation velocity of \texttt{HI} (and also \texttt{CO}) should be reflected in Doppler-measurements of the 511\,keV line in high resolution if these regions are indeed the annihilation sites for most of the positrons.
	While this is not excluded, \citet{Siegert2019_lv511} find that the rotation curve as measured in 511\,keV is also consistent with zero.
	Also here, the annihilation-in-flight continuum would be expected.

	\subsection{Inverse Compton Scattering}
	
	Inverse Compton (\texttt{IC}, Fig.\,\ref{fig:IC_map}) scattering appears from reactions of charged particles with the ambient photon fields, mainly from the CMB or stellar light.
	Here, cosmic rays, as probably accelerated by supernovae and their remnants, compact objects or stellar winds, propagate through the ISM and emit at X- and $\gamma$-ray energies via the production of secondary particles or \texttt{IC} interactions.
	At MeV energies, the diffuse continuum $\gamma$-ray spectrum is dominated by \texttt{IC} scattering of cosmic-ray electrons (and positrons) with the stellar radiation field.
	It directly shows where the positrons are expected to propagate and loose energy towards their final annihilation.
	Thus also \texttt{IC} scattering templates may show the expected annihilation in flight continuum.
	For positrons, the propagation through a typical ISM may take up to several $10^5\,\mrm{yrs}$ on average, thereby losing energy via Coulomb interactions, synchrotron radiation, \texttt{IC}, and bremsstrahlung.
	Afterwards, the positrons can thermalise with the ambient medium and remain there on time scales of several $10^{5}\,\mrm{yrs}$ until they annihilate.
	However, if the positrons find an annihilation site close to their sources, the annihilation may also be directly traced by a \texttt{IC} emission template, as energy loss and annihilation would be spatially close.
	
	% \begin{table}[!ht]
	%   \centering
	%       \begin{tabular}{ c c c c c c  }\hline \hline
	%       $R_h$ & $z_h$ & CR source distribution  & $E(B-V)$ cut & $T_s$\\ 
	%       (kpc) & (kpc) & & (mag) &(K) \\ \hline 
	%          20 & 10 &  Lorimer & 2 &150 \\ \hline \hline
	%     \end{tabular}
	%     \caption{\label{tab:galprop_setup} Main GALPROP propagation
	%     parameter setup considered in this study. Other propagation parameters (that minimally impact the resulting maps) can be obtained from ref.\,\citep{Ackermann2012_FermiLATGeV}.}
	% \end{table} 
	
	The propagation of positrons in the Galaxy can be studied with various publicly available tools \cite[e.g., \texttt{GALPROP} or \texttt{DRAGON},][]{Strong2007_GALPROP,Evoli2008_CRdiffuse}.
	In particular, \texttt{GALPROP} contains a set of routines that solve the particle transport equation via sophisticated numerical methods.
	Given a certain Galactic structure (e.g. interstellar gas, radiation and magnetic fields), cosmic-ray source distribution, injection spectrum and boundary conditions, \texttt{GALPROP} makes detailed predictions of observable signature for all cosmic-ray species.
	Simulations performed with this package can include pure diffusion, diffusive re-acceleration (diffusion in energy space), convection (Galactic winds),  energy losses (ionisation, Coulomb scattering, bremmstrahlung, IC scattering, and synchrotron radiation), nuclear fragmentation, and radioactive decay \citep{Moskalenko2005_CRprop}.
	
	\citet{Ackermann2012_FermiLATGeV} performed maximum-likelihood fits to all-sky Fermi-LAT gamma-ray data using a grid of alternative GALPROP models constructed using different propagation parameter setups.
	The study found that the set of parameters that had the largest impact in the fits were cosmic-ray source distribution (supernova remnants or pulsars), cosmic-ray halo height $z_h$, cosmic-ray halo radius $R_h$, the atomic hydrogen spin temperature ($T_s$) and the magnitude cut applied to the $E(B-V)$ reddening maps used in the construction of the dust maps.
	Here we use GALPROP v54 \citep{Strong2007_GALPROP} to create energy-dependent \texttt{IC} emission templates for the sky regions covered in our data set.
	We assume one of the propagation parameter setups considered in \citep{Ackermann2012_FermiLATGeV} for our standard IC model map, i.e. scale radius $R_h = 20$\,kpc, scale height $z_h = 10$\,kpc, a Lorimer cosmic-ray source distribution, an $E(B-V)$ cut of $2$\,mag, and a temperature $T_s = 150$\,K.
	
	In Fig.\,\ref{fig:IC_map}, we show the resulting \texttt{IC} map between 245 and 319\,keV.
	With increasing energy, the apparent scale height as well as the scale radius decrease in the sky region $|l|\leq25^{\circ}$, $|b|\leq25^{\circ}$.
	While between 189 and 245\,keV, the scale height is about $9.8^{\circ}$, the line-of-sight integration yields only $8.9^{\circ}$ between 800 and 1805\,keV.
	This is explained by the anisotropy of the radiation field that influences the \texttt{IC}-spectrum as a function of energy \citep{Moskalenko2000_IC}.
	The \texttt{IC} emission templates in these energy bands are slightly asymmetric towards positive longitudes with maxima around $l=0.7^{\circ}$.
	\texttt{IC} emission is expected in all energy bands, so that \texttt{IC} templates may serve as baseline models to asses different positron-annihilation templates (cf. Sec.\,\ref{sec:baseline_model}).

	\subsection{Dark Matter Density Profile}
	
	\citet{Skinner2014_511} showed that a dark matter density profile, taken to the power of two and then line-of-sight integrated fits the SPI 511\,keV data.
	This could be tentatively interpreted as electron-positron pairs from dark matter annihilation -- as a result from a two-body process.
	However, this would only be case if the positrons annihilate close to their production sites, as described by a dark matter halo.
	Here we use the general NFW profile \citep{Navarro1997_NFW}
	
	\begin{equation}
		\rho_{\rm DM}(x,y,z) = \frac{\rho_0}{(R/R_0)^{\gamma}\left[1+(R/R_0)^{\alpha}\right]^{(\beta-\gamma)/\alpha}}\mrm{,}
		\label{eq:NFW_profile}
	\end{equation}
	
	\noindent with $\alpha=1$, $\beta=3$, and $R_0=20\,\mrm{kpc}$ \citep{Abazajian2014_DMGeV}, as was shown to fit both the GeV \citep{Hooper2011_DMGeV,Abazajian2014_DMGeV,Daylan2016_GeVDM} as well as the 511\,keV data \citep{Vincent2012_dm511,Skinner2014_511}.
	The inner slope $\gamma$ has been estimated from GeV studies to be around $1.2$ and classically was introduced as $1.0$.
	Also, \citet{Skinner2014_511} used a profile with $\gamma=1.0$ to compare directly to SPI data.
	For this reason, we introduce two halo templates, one with $\gamma$-factor of $1.0$, which we call \texttt{DM0}, and one with $\gamma=1.2$, called \texttt{DM2}.
	These profiles may also be tracers for the stellar halo population, and may not be uniquely identified as due to dark matter annihilation.
	The latter would describe WIMP-like particle self-annihilations into electron-positron pairs, which in turn might quickly annihilate to 511\,keV photons, either directly or by the formation of Positronium.
	
	The templates, Fig.\,\ref{fig:DM0_map}, are axisymmetric around $(l,b)=(0^{\circ},0^{\circ})$ with a pronounced peak.
	In our region of interest, 99\% of the expected emission from \texttt{DM0} (\texttt{DM2}) is included within a circle of $9^{\circ}$ ($2^{\circ}$), so that \texttt{DM0} covers the size of the Galactic bulge and \texttt{DM2} the size of the nuclear bulge.
	Note that due to the angular resolution of SPI, the templates \texttt{DM0} and \texttt{DM2} are very similar to the inner bulge template \texttt{NB}, and the differences are mostly in the wings of the distributions.
	We exclude combinations of the templates \texttt{DM0} and \texttt{DM2} in the analysis as we want to trace individual populations and two dark matter profiles in combination would be unphysical.

	\subsection{Fermi Bubbles}
	
	The Fermi Bubbles (\texttt{FB}) have first been identified by \citet{Su2010_fermibubbles} as giant bipolar GeV emission at high latitudes above the Galactic plane.
	The emission spectrum ($\propto E^{-2}$) differs from IC scattering and pion-production.
	Different scenarios have been suggested to explain this emission feature, from past activity of an active Galactic nucleus of the Milky Way \citep{Su2010_fermibubbles,Zubovas2012_FermiBubbles_AGN,Guo2012_FermiBubbles_AGN,Yang2017_FermiBubbles_AGN}, a Galactic wind, jet, or outflow \citep{Crocker2012_FermiBubbles_outflows,Sarkar2015_FermiBubbles_outflows,Lacki2014_FermiBubbles_outflows}, connections to more local structures such as Loop I or the North Polar Spur \citep{Kataoka2013_FermiBubbles_local,Sofue2015_NPS_Aquila_FermiBubbles,Ackermann2014_FermiBubbles}, or the WMAP haze \citep{Dobler2012_FermiBubbles_WMAPHaze,Ackermann2014_FermiBubbles,Crocker2015_FermiBubbles_WMAPHaze}.
	All these emission mechanisms could be accompanied by positron production so that the same morphology as seen in GeV could be prominent in positron annihilation.
	This includes direct annihilation in flight for the high-energy continuum as well as the formation of Positronium for the 511\,keV line and the annihilation continuum below.
	
	We use a flat, i.e. isotropically filled, template as was suggested by \citet{Macias2018_LATGeV} for the GeV emission.
	The bubbles write
	
	\begin{equation}
		F_{\rm FB}(l,b) \propto l_0 \times \left( \cosh((l-l_1)/l_0) - l_2 \right)\mrm{,}
		\label{eq:FB_formula}
	\end{equation}
	
	\noindent where $l_0 = 10.5^{\circ}$ ($l_0 = 8.7^{\circ}$), $l_1 = 1.0^{\circ}$ ($l_1 = -1.7^{\circ}$), and $l_2 = 1.0^{\circ}$ ($l_2 = 1.0^{\circ}$) for the northern (southern) bubble \citep{Acero2016_FermiLAT_sources}.
	The different scenarios for the Fermi Bubbles also try to explain the asymmetry in the template, mainly due to line-of-sight effects or inclinations angles.
	In general, the \texttt{FB}-template, Fig.\,\ref{fig:FB_map}, traces morphologies as would be expected from out- or inflows from or to the Galactic centre.
	Also here, annihilation in flight could naturally be expected from the kinematics of the explanatory scenarios.
	We assume isotropy in the \texttt{FB} template as the granularity of the GeV emission is hard to identify \citep{Ackermann2014_FermiBubbles}.
	The bubbles extend to latitudes $|b| \lesssim 45^{\circ}$.

	\section{Results for Classical ISM Phases}\label{sec:ISM_phases_results}
	In our assessment of the propagation distance spread and therefore injection energy of positrons, we used a parametrised grid of ISM conditions in density and ionisation fraction as well as injection energies.
	We assume a complete random walk after each collision which acknowledges the uncertain magnetic field structure in the bulge so that also back-scatters are allowed.
	This is the limiting case of least possible propagation in a certain medium, which is equivalent to the highest possible injection energy.

\begin{figure}
		\centering
		\includegraphics[width=0.8\columnwidth,trim=0.10in 0.1in 0.6in 0.6in,clip=true]{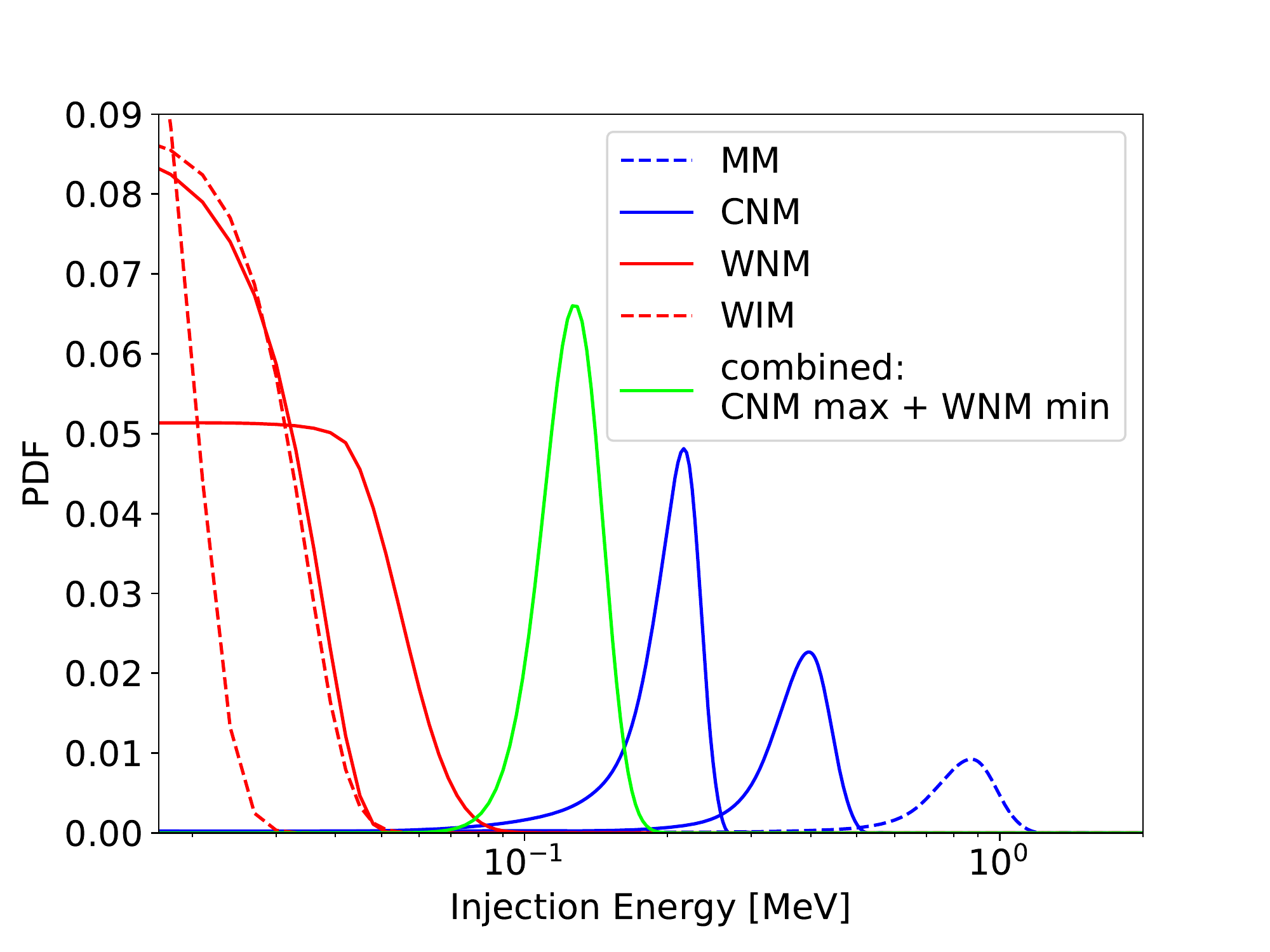}%
		\caption{Similar to Fig.\,\ref{fig:limits_Einj} but for canonical ISM phases as described in \citet{Jean2009_511ISM}. The two cases for each line style represent the minimum and maximum values considering the density and ionisation fraction.}%
		\label{fig:ISM_Einj_results_appendix}%
	\end{figure}

	\citet{Jean2009_511ISM} performed a detailed 3D propagation of MeV positrons, taking into account pitch angle scattering.
	In their work, they present the results in classical ISM phases with realistic densities and ionisation fraction.
	In particular they show propagation distances in the molecular medium (MM; $n_{\rm H} = 10^2$--$10^6\,\mrm{cm^{-3}}$, $x_{\rm ion} \lesssim 10^{-4}$), cold neutral medium (CNM; $n_{\rm H} = 20$--$50\,\mrm{cm^{-3}}$, $x_{\rm ion} \lesssim 10^{-3}$), warm neutral medium (WNM; $n_{\rm H} = 0.2$--$0.5\,\mrm{cm^{-3}}$, $x_{\rm ion} \lesssim 0.05$), warm ionised medium (WIM; $n_{\rm H} = 0.2$--$0.5\,\mrm{cm^{-3}}$, $x_{\rm ion} = 0.6$--$0.9$), and hot ionised medium (HIM; $n_{\rm H} = 0.005$--$0.01\,\mrm{cm^{-3}}$, $x_{\rm ion} = 1$) (cf. their Table 1).
	We use these minimum and maximum values to infer bounds on the injection energy given our distance spread measurements in NB and BB in Sec.\,\ref{sec:results}.
	In Fig.\,\ref{fig:ISM_Einj_results_appendix}, we show the resulting bounds on the injection energy in NB, BB, and a combined analysis.
	There is no significant difference in the general conclusion that the NB and BB hardly match in terms injection energy and thus of positron sources.
	Also similar in this case is the minimal $3\sigma$ overlap in the two bounding cases of the CNM and WNM, however now at smaller injection energies.
	Using the canonical ISM phases rather than measured densities in NB and BB results in an allowed range of injection energies of $0.13 \pm 0.02$\,MeV.
	This rather small allowed range might be too optimistic because it is known that the NB has densities about ten times higher than the canonical CNM, and also the BB might show larger densities than only $0.5\,\mrm{cm^{-3}}$ as in the WNM or WIM.
	Extrapolating the values from \citet{Jean2009_511ISM} to these larger densities results in a difference between the two cases of inefficient pitch angle scattering and complete random walk of a factor four.
	We suggest to use this value as a measure of the systematic uncertainty of this propagation distance analysis.
	Therefore, taking into account both scenarios, we find that if the same sources are responsible in the NB and BB, they inject positrons at energies of at most $0.4$--$1.4$\,MeV.

	\section{Data Set Details}\label{sec:data_set_details}
	
	Our data set comprises INTEGRAL/SPI observations (pointings) inside the spherical rectangle region $\Delta l \times \Delta b = 15^{\circ} \times 15^{\circ}$ around the Galactic centre, $(l,b) = (0^{\circ},0^{\circ})$.
	The fully-coded field of view of SPI is $16^{\circ}\times16^{\circ}$, and the partially-coded field of view $30^{\circ}\times30^{\circ}$, i.e. not all detectors are exposed to the source.
	Because of SPI's large field of view, the analysis region is extended by at least $16^{\circ}/2 = 8^{\circ}$ in all directions.
	This is taken into account when spatial templates are created and fitted, and also when MeV flux estimates are compared/extrapolated to GeV energies, for instance.
	The exposure map for this data set is shown in Fig.\,\ref{fig:expo_map}.
	The definition of the energy bins is listed in Tab.\,\ref{tab:data_set_energies}, together with additional characteristics concerning instrumental background and potential emission mechanisms.
	
	\begin{table*}
		\centering
		\caption{Summary of energy bins, number of data points ($N_{\rm obs}$), and background (BG) scaling used in this analysis (cf. Sec.\,\ref{sec:likelihood_analysis}), resulting in $N_{dof}$ degrees of freedom. The last column describes the expected dominant contributions at these energies, according to \citet{Strong2005_gammaconti} and this work. The resolved point sources (res. PSs), in the first two energy bins are 1E 1740.7-2942, GRO J1655-40, GRS 1758-258, IGR J17464-3212, IGR J17475-2822, and SLX 1744-299. Pulse-shape discrimination (PSD) data has been used to account for SPI's `electronic noise' region.}
		\begin{tabular}{crrcc}
			\hline
			\hline
			Energy band [keV] & $N_{\rm obs}$ & $N_{\rm dof}$ & BG scaling & Phys. mechanisms \& Comment \\
			\hline
			189--245  & 203138 & 193531 & $\sim 1.5$~h & IC, unres. PSs, 6 res. PSs, time-dep. \\
			245--319  & 203138 & 194112 & $\sim 1.5$~h & IC, unres. PSs, 1 res. PS, $\mrm{e^+e^-}$-oPs \\
			319--414  & 203138 & 194113 & $\sim 1.5$~h & IC, unres. PSs, $\mrm{e^+e^-}$-oPs \\
			414--508  & 203138 & 190063 & $\sim 1.0$~h & IC, $\mrm{e^+e^-}$-oPs \\
			508--514  & 203138 & 202515 & $\sim 6.0$~d & IC, $\mrm{e^+e^-}$, 511\,keV line\\
			514--800  & 202480 & 193706 & $\sim 1.5$~h & IC, PSD data selection \\
			800--1805 & 202480 & 193706 & $\sim 1.5$~h & IC, PSD data selection \\
			\hline
			\hline
		\end{tabular}
		\label{tab:data_set_energies}
	\end{table*}

	We want to note that the analysis procedure for SPI is fundamentally different, compared to Fermi/LAT, for example, in which the measured counts are projected onto a sky map, and the resulting image is interpreted in terms of background and different templates.
	On the other hand, it is very difficult to provide morphology-independent spectra or spectrum-independent morphologies with SPI, as the count rate is dominated by instrumental background.
	A projection of SPI events onto a celestial sphere will show instrumental background and is difficult to be analysed in this dimension.
	Consequently, the data analysis takes place in the raw detector$\times$pointing format.

	%%%%%%%%%%%%%%%%%%%%%%%%%%%%%%%%%%%%%%%%%%%%%%%%%%

	% Don't change these lines
	\bsp	% typesetting comment
	\label{lastpage}
\end{document}